\documentclass[12pt]{article}
\usepackage{newinutile}
\usepackage{amsmath,amssymb}
\usepackage{epsfig,graphics,color,calc,graphicx}
\usepackage{cite}
\usepackage{mathrsfs}
\usepackage{wasysym}
\usepackage{url}

\newcommand{\newatop}[2]{\genfrac{}{}{0pt}{}{#1}{#2}}

\newcommand{\rr}[1]{{\normalfont\textrm{#1}}}
\newcommand{\cc}[1]{{\mathcal{#1}}}
\newcommand{\bb}[1]{{\mathbb{#1}}}

\newcommand{\puno}{{\textbf{u}}}
\newcommand{\muno}{\mathbf{d}}
\newcommand{\chess}{\mathbf{c}}

\newlength{\pecettawidth}
\begin{document}
\title{Effect of self--interaction on the phase diagram  of 
a Gibbs--like measure derived by a 
reversible Probabilistic Cellular Automata}

\author{Emilio N.M.\ Cirillo}
\email{emilio.cirillo@uniroma1.it}
\affiliation{Dipartimento di Scienze di Base e Applicate per 
             l'Ingegneria, Sapienza Universit\`a di Roma,
             via A.\ Scarpa 16, I--00161, Roma, Italy.}

\author{Pierre--Yves Louis}
\email{pierre-yves.louis@math.univ-poitiers.fr}
\affiliation{Laboratoire de Math\'ematiques et Applications, UMR 7348 Universit\'e de Poitiers \& CNRS, T\'el\'eport 2 - BP 30179
Boulevard Marie et Pierre Curie, F--86962 Technopole du Futuroscope de Poitiers Cedex, France}

\author{Wioletta M.\ Ruszel}
\email{W.M.Ruszel@tudelft.nl}
\affiliation{Delft Institute of Applied Sciences, 
Technical University Delft,
Mekelweg 4, 2628 CD Delft, The~Netherlands}

\author{Cristian Spitoni}
\email{C.Spitoni@uu.nl}
\affiliation{Institute of Mathematics, 
University of Utrecht, Budapestlaan 6, 3584 CD Utrecht, The~Netherlands}


\begin{abstract}
Cellular Automata are discrete--time dynamical systems on a 
spatially extended discrete space which 
provide paradigmatic examples of 
nonlinear phenomena.
Their stochastic generalizations, i.e., 
Probabilistic Cellular Automata (PCA), 
are discrete time Markov chains
on lattice with finite single--cell states whose 
distinguishing feature is the \emph{parallel} character of the updating rule.
We study the ground states of the 
Hamiltonian and the low--temperature phase diagram of the related 
Gibbs measure
naturally associated with a   
class of reversible PCA, called the \textit{cross PCA}. 
In such a model the updating rule of a cell 
depends indeed only on the status of the five cells forming a cross 
centered at the original cell itself. 
In particular, it depends on the value of the center spin 
(\textit{self--interaction}). 
The goal of the paper is that of investigating the 
role played by the self--interaction parameter
in connection with the ground states of the Hamiltonian and the 
low--temperature phase diagram of the Gibbs measure associated 
with this particular PCA.
\end{abstract}





\maketitle

\section{Introduction}
\label{s:introduzione}
\par\noindent
Cellular Automata (CA) are discrete--time dynamical systems on a 
spatially extended discrete space. 
They are well known for -- at the same time -- being easy to define 
and implement and
for exhibiting a rich and complex nonlinear behavior as 
emphasized for instance in~\cite{Wolfram1984,Wolfram1984Nature} 
for CA on one--dimensional lattice. See~\cite{NonLinearBook} to precise
the connections with the nonlinear physics.
For the general theory of deterministic CA we refer to the recent
paper~\cite{Kari} and references therein.

Probabilistic Cellular Automata (PCA) are CA straightforward generalization 
where the updating rule is stochastic.
They inherit the computational power of CA and are used as models in 
a wide range of applications (see,
for instance, the contributions in~\cite{sirakoulis2012cellular}).
From a theoretic perspective, the main challenges 
concern the non--ergodicity of these dynamics for an infinite collection
of interacting cells. 
Ergodicity means the non--dependence of the long--time behavior on the initial 
probability distribution and the convergence in law towards a unique 
stationary probability distribution (see~\cite{tvs} for details and 
references).
Non--ergodicity is related to \emph{critical phenomena} and it 
is sometimes referred to as \emph{dynamical phase transition}.

Strong relations exist between PCA and the general equilibrium statistical 
mechanics framework~\cite{Wolfram:RevModPhys.55.601,GeorgesLeDoussal,lms}. 
Important issues are related to the interplay between disordered global states 
and ordered phases (\emph{emergence of organized global states, phase 
transition})~\cite{PCA:order:disorder}. 
Altough, PCA initial interest arose in the framework of 
Statistical Physics, in the recent literature many different applications
of PCA have been proposed. In particular it is notable to remark 
that a natural context in which the PCA main ideas are of interest 
is that of evolutionary games \cite{PGSFM,PG,PS}.

PCA dynamics are naturally defined 
on an infinite lattice.
Given a local stochastic updating rule, one has to face the usual 
problems about the connections 
between the PCA dynamics on a finite subpart of the lattice
and the dynamics on
the infinite lattice. 
In particular, it was stated in~\cite{gklm} for translation--invariant 
infinite volume PCA with \emph{positive rates}\footnote{A 
PCA is said to be with \emph{positive rates} if the local updating rule
is a distribution giving positive probability to any cell--state.}, 
that the law of the trajectories, starting from any stationary 
translation--invariant distribution, 
is the Boltzmann--Gibbs distribution   
for some space--time associated potential.
Thus phase transition for the space--time potential is intimately 
related to the PCA dynamical phase transition.

Moreover, see \cite[Proposition~2.2]{DpLR}, 
given a translation--invariant PCA dynamics, if there exists one
translation--invariant stationary distribution which is a 
Gibbs measure with respect to some potential on the lattice, 
then all the associated translation--invariant stationary distributions 
are Gibbs with respect to the the same potential.

In this paper we shall consider a particular class of PCA, called 
\textit{reversible} PCA, which are reversible with respect to 
a Gibbs--like measure defined via a translation invariant 
multi--body potential. In this framework we shall study the 
zero and low--temperature 
phase diagram of such an equilibrium statistical mechanics--like system, 
whose phases are related to the stationary measures of the 
original PCA. 

We shall now first briefly recall formally the 
definitions of Cellular Automata and 
Probabilistic Cellular Automata and then describe the main results
of the paper.

\subsection{Cellular Automata}
\label{s:ca}
\par\noindent
Cellular Automata are defined via a local deterministic evolution 
rule. 
Let $\Lambda\subset\bb{Z}^d$ be a finite cube with periodic boundary 
conditions.

Associate with each site $i\in\Lambda$ (also called \textit{cell})
the \text{state variable}
$\sigma_i\in\mathcal{S}_0$, where $\mathcal{S}_0$ is a finite single-site space and denote by 
$\Omega:=\mathcal{S}_0^\Lambda$ the \textit{state space}.
Any $\sigma\in\Omega$ is called a \textit{state} or \textit{configuration}
of the system. 

In order to define the evolution rule we consider 
$I$, a subset of the torus $\Lambda$, and 
a function 
$f_I:\mathcal{S}_0^I \to\mathcal{S}_0$ 
depending on the state variables in $I$.
We also introduce the shift $\Theta_i$ on the torus, for any $i\in\Lambda$, 
defined as the map 
$\Theta_i:\Omega\to\Omega$ 
\begin{equation}
\label{shift}
(\Theta_i\sigma)_j=\sigma_{i+j}.
\end{equation}
The configuration $\sigma$ at site $j$ shifted by $i$ is equal to the configuration at site $i+j$. 
For example (see figure~\ref{f:def}) set $j=0$, then the value of the spin at the origin $0$ will be mapped to site $i$. 
The \textit{Cellular Automaton} on $\Omega$ with rule $f_I$ is the 
sequence 
$\sigma(0),\sigma(1),\dots,\sigma(t)$, for $t$ a positive integer,
of states in $\Omega$ satisfying the following (deterministic) rule:
\begin{equation}
\label{regola-det}
\sigma_i(t)=f_I(\Theta_{i}\sigma(t-1)) 
\end{equation}
for all $i\in\Lambda$ and $t\ge1$.

Note the local and parallel character of the evolution:
the value $\sigma_i(t+1)$, 
for all $i\in\Lambda$,
of all the state variables at time $t+1$ depend on the 
value of the state variables at time $t$ (parallel evolution) 
associated only with the sites in $i+I$ (locality). 

\begin{figure}[t]
\setlength{\unitlength}{1.5pt}
 \begin{picture}(10,55)(-150,-5)
 \thinlines
 \multiput(-5,0)(0,10){5}{\put(0,0){\line(1,0){50}}}
 \multiput(0,-5)(10,0){5}{\put(0,0){\line(0,1){50}}}
 \put(-12,20){${\scriptstyle \Lambda}$}
 \put(10,20){\circle*{2}}
 \put(11,21){${\scriptstyle 0}$}
 \thicklines
 \put(5,15){\line(0,1){20}}
 \put(5,15){\line(1,0){10}}
 \put(5,35){\line(1,0){20}}
 \put(15,15){\line(0,1){10}}
 \put(15,25){\line(1,0){10}}
 \put(25,25){\line(0,1){10}}
 \put(1,13){${\scriptstyle I}$}
 \put(30,0){\circle*{2}}
 \put(31,1){${\scriptstyle i}$}
 \put(25,-5){\line(0,1){20}}
 \put(25,-5){\line(1,0){10}}
 \put(25,15){\line(1,0){20}}
 \put(35,-5){\line(0,1){10}}
 \put(35,5){\line(1,0){10}}
 \put(45,5){\line(0,1){10}}
 \put(47,14){${\scriptstyle i+I}$}
 \end{picture}
 \caption{Schematic representation of the action of the 
          shift $\Theta_{i}$ defined in \eqref{shift}.}
 \label{f:def}
 \end{figure}

\subsection{Probabilistic Cellular Automata}
\label{s:pca}
\par\noindent
The stochastic version of Cellular Automata is called 
\textit{Probabilistic Cellular Automata} (PCA).
We consider a
probability distribution
$f_{\sigma}:\mathcal{S}_0\to[0,1]$ depending
on the state $\sigma$ restricted to $I$; we drop the dependence on $I$ in the notation for future convenience.
A Probabilistic Cellular Automata is 
the Markov chain $\sigma(0),\sigma(1),\dots,\sigma(t)$
on $\Omega$ with transition matrix
\begin{equation}
\label{regola-pca}
p(\sigma,\eta)=\prod_{i\in\Lambda}f_{\Theta_{i}\sigma} (\eta_i)
\end{equation}
for $\sigma,\eta\in\Omega$. We remark that  $f$ depends on $\Theta_{i}\sigma$ only via the neighborhood $i+I$. 
Note that, as in the deterministic case, the character of the 
evolution is local and parallel.

\subsection{Description of the problem and results}
\label{s:syn}
\par\noindent
Under suitable hypotheses on the probability distribution $f_{\sigma}$, for 
$\Lambda$ finite, 
the Markov chain is irreducible and aperiodic, so that a unique 
stationary probability measure exists. On the other hand, 
irreducible and aperiodic PCA
are in general 
not reversible.
As already proven in 
\cite{KV,tvs,GJH} there exists a class of 
PCA which are reversible with respect to 
a Gibbs--like probability measure \cite[Proposition~3.1]{DpLR} and,
hence, they admit a sort of Hamiltonian. 
These models will be called \textit{reversible PCA} (see 
\cite[Section~3.5]{ThesePyl} for more details).

From the results in \cite{DpLR}, see for instance 
Proposition~3.3 therein, it is possible to deduce that 
these Gibbs--like measures are either 
stationary or two--periodic for the PCA.
Therefore it is quite natural 
to compare the behavior of these distributions to the one of the
statistical mechanics counterpart. 

Moreover, it is worth mentioning that also non--equilibrium 
properties of the PCA dynamics 
have been widely investigated. 
In~\cite{EspTempsPyl}, 
in the attractive reversible case and in absence of phase transition, 
the equivalence
between an equilibrium weak--mixing condition and the 
convergence towards a unique equilibrium state with exponential speed
was proven.
In~\cite{BCLS,CN,CNS1,CNS2,NS} the  
metastable behavior of a certain class of reversible PCA has been analyzed.
In this framework the remarkable 
interest of a particular reversible PCA has been pointed out, called the 
cross PCA (see Section~\ref{s:croce}). It is a two--dimensional 
reversible PCA in which the updating rule of a cell 
depends on the status of the five cells forming a cross 
centered at the cell itself. 
In this model, the future state of the spin at a given cell 
depends also on the present value of such a spin. This 
effect will be called \textit{self--interaction} and its weight 
in the updating rule 
will be called \textit{self--interaction intensity}.

In~\cite{CNS3} the analogies between the metastable behavior 
of the cross PCA and the Blume--Capel model~\cite{B,BEG,C} have been 
pointed out. Starting from~\cite{CNS1}, 
 it has been heuristically argued that from a metastability 
perspective, the cross model behaves like the Blume--Capel 
one once the checkerboard configuration and the self--interaction intensity 
of the cross model are identified with the empty configuration 
and the chemical potential of the Blume--Capel one~\cite{CO}. 

In this paper we shall investigate this analogy further from an 
equilibrium point of view. 
We study the low--temperature\footnote{Note that 
in the case of reversible PCA the use of the word 
temperature is misleading 
since the stationary measure is not precisely a Gibbs one. 
But, as it will be discussed in Section~\ref{s:rpca}, a parameter playing 
a similar role can be introduced.}
phase diagram of the Gibbs--like measure associated with the cross PCA
which, as explained above, is strictly connected to the structure of the 
stationary states of the PCA. 
This is a very difficult task, since the microscopic 
interaction is described by a Hamiltonian in which 
coupling constants associated with 
all the multi--body potentials that can be constructed 
inside a five site cross are present. 
As a first step we shall discuss the zero--temperature 
phase diagram, namely, the structure of the ground states 
of the system, and we will show that the analogy with the Blume--Capel 
model is still strict. 
The second step will be the study of how the phase diagram changes 
when the temperature is fixed to a small positive 
value. In this case we will see that great differences, at least at the 
level of the Mean Field approximation, between the 
Blume--Capel and the cross PCA case will emerge. 

One of the distinguishing features of the Blume--Capel model 
is the presence of a triple point in the zero--temperature phase 
diagram corresponding to zero chemical potential and magnetic field. 
In this point the three ground states (homogeneous plus, minus, and 
lacuna state) coexist (see, for instance, \cite[Fig.~1]{S}). 
It was proven in~\cite{S,BS} that this triple 
point moves toward the region with positive chemical potential 
when the temperature is positive and small. This is an entropy 
effect explained in~\cite{S}.

In the present paper for the cross PCA model we prove a similar 
structure for the zero--temperature 
phase diagram: the triple point is at zero self--interaction intensity and 
magnetic field. At this point the four ground states (homogeneous plus, 
minus, even checkerboard, and odd checkerboard) coexist. Due to the 
presence of four coexisting ground states, an entropy argument similar 
to the one developed for the Blume--Capel model suggests that the position of 
the triple point is not affected by a small positive 
temperature~\cite[Section~2.4]{CN}. 
In this paper we approach the problem also from a Mean Field point of 
view and we obtain a result consistent with this conjecture. 

The paper is organized as follows. 
In Section~\ref{s:rpca-sezione} we introduce the 
reversible Probabilistic Cellular Automata and discuss some general 
properties. 
In Section~\ref{s:croce} we introduce the cross PCA and discuss 
its Hamiltonian. In particular we study its ground states and 
draw the zero temperature phase diagram. 
In Section~\ref{s:mf} we study the phase diagram 
of the cross PCA in the framework of the Mean Field approximation. 
Finally, we summarize our conclusions in Section~\ref{s:conclusioni}.
The technical details of the painstaking computations we had to perform
have been relegated to the Appendix. 

\section{Reversible Probabilistic Cellular Automata}
\label{s:rpca-sezione}
\par\noindent
In this section we shall introduce reversible Probabilistic Cellular Automata
and discuss some general results on their Hamiltonian. 

\subsection{Reversible Probabilistic Cellular Automata}
\label{s:rpca}
\par\noindent
A class of \textit{reversible} PCA can be obtained by choosing 
$\Omega=\{-1,+1\}^\Lambda$,
and
\begin{equation}
\label{frev}
 f_{\sigma;h}(s)=\frac{1}{2}\Big\{
        1+s\tanh\Big[\beta\Big(\sum_{j\in\Lambda}k(j)\sigma_j+h\Big)\Big]\Big\}
\end{equation}
for all $s\in\{-1,+1\}$
where $T\equiv1/\beta>0$ 
and $h\in\mathbb{R}$ are called \textit{temperature} and 
\textit{magnetic field}. $k:\bb{Z}^2\to\bb{R}$ is such that its 
support is a subset of~$\Lambda$ and~$k(j)=k(j')$
whenever $j,j'\in\Lambda$ are symmetric with respect to the origin. 
Recall that, by definition, the support of the function $k$ is the subset 
of $\Lambda$ where the function~$k$ is different from zero.
With the notation introduced above, the set~$I$ is the support 
of the function~$k$.

Recall that $\Lambda$ is a finite torus, namely, periodic boundary 
conditions are considered throughout this paper. 
It is not difficult to prove~\cite{GJH,KV}
that the above specified PCA dynamics is reversible 
with respect to the finite--volume Gibbs--like 
measure\footnote{This statement, with some care,
can be extended to non--periodic 
boundaries \cite{DpLR}. 
For finite lattice and periodic boundary conditions 
the finite--volume Gibbs distribution is the unique
reversible one (our framework).
For finite lattice and fixed deterministic non--periodic 
boundary conditions, the finite--volume Gibbs distribution differs from
the unique reversible one; differences are somehow 
localized close to the boundary .}
\begin{equation}
\label{gibbs}
 \mu_{\beta,h}(\sigma)
 =
 \frac{1}{Z_\beta}
\,e^{-\beta G_{\beta,h}(\sigma)}
\end{equation}
with \textit{Hamiltonian}
\begin{equation}
\label{mahG}
G_{\beta,h}(\sigma)
=
 -h\sum_{i\in\Lambda}\sigma_i
 -\frac{1}{\beta}\sum_{i\in\Lambda}
    \log\cosh\Big[\beta
   \Big(\sum_{j\in\Lambda}k(j-i)\sigma_j+h\Big)\Big]
\end{equation}
and \textit{partition function}
\begin{equation}
\label{partition}
 Z_{\beta,h}=
 \sum_{\eta\in\Omega} e^{-\beta G_{\beta,h}(\eta)}
\end{equation}
In other words, in this case the detailed balance equation
\begin{displaymath}
p(\sigma,\eta) e^{-\beta G_{\beta,h}(\sigma)}
=e^{-\beta G_{\beta,h}(\eta)} p(\eta,\sigma)
\end{displaymath}
is satisfied thus the probability measure~$\mu_{\beta,h}$ is stationary for 
the PCA. 

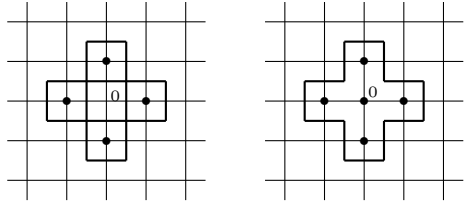
\begin{figure}[t]
\setlength{\unitlength}{1.5pt}
 \begin{picture}(10,55)(-115,-5)
 \thinlines
 \multiput(-5,0)(0,10){5}{\put(0,0){\line(1,0){50}}}
 \multiput(0,-5)(10,0){5}{\put(0,0){\line(0,1){50}}}
 \put(20,30){\circle*{2}}
 \put(30,20){\circle*{2}}
 \put(20,10){\circle*{2}}
 \put(10,20){\circle*{2}}
 \put(21,20){${\scriptscriptstyle 0}$}
 \thicklines
 \put(5,15){\line(0,1){10}}
 \put(15,15){\line(0,1){10}}
 \put(25,15){\line(0,1){10}}
 \put(15,15){\line(1,0){10}}
 \put(5,25){\line(1,0){10}}
 \put(15,25){\line(0,1){10}}
 \put(15,25){\line(1,0){10}}
 \put(15,35){\line(1,0){10}}
 \put(25,35){\line(0,-1){10}}
 \put(25,25){\line(1,0){10}}
 \put(35,25){\line(0,-1){10}}
 \put(35,15){\line(-1,0){10}}
 \put(25,15){\line(0,-1){10}}
 \put(25,5){\line(-1,0){10}}
 \put(15,5){\line(0,1){10}}
 \put(15,15){\line(-1,0){10}}

 \thinlines
 \multiput(60,0)(0,10){5}{\put(0,0){\line(1,0){50}}}
 \multiput(65,-5)(10,0){5}{\put(0,0){\line(0,1){50}}}
 \put(85,30){\circle*{2}}
 \put(95,20){\circle*{2}}
 \put(85,10){\circle*{2}}
 \put(75,20){\circle*{2}}
 \put(85,20){\circle*{2}}
 \put(86,21){${\scriptscriptstyle 0}$}
 \thicklines
 \put(70,15){\line(0,1){10}}
 \put(70,25){\line(1,0){10}}
 \put(80,25){\line(0,1){10}}
 \put(80,35){\line(1,0){10}}
 \put(90,35){\line(0,-1){10}}
 \put(90,25){\line(1,0){10}}
 \put(100,25){\line(0,-1){10}}
 \put(100,15){\line(-1,0){10}}
 \put(90,15){\line(0,-1){10}}
 \put(90,5){\line(-1,0){10}}
 \put(80,5){\line(0,1){10}}
 \put(80,15){\line(-1,0){10}}
 \end{picture}
\caption{Schematic representation of the nearest neighbor (left) 
         and cross (right) models.}
\label{f:modelli}
\end{figure}

Note that different reversible PCA models can be specified by 
choosing different functions~$k$. In particular the support~$I$ of such 
a function can be varied. The generality of this PCA family among the reversible one 
was remarked in Section~4.1.1 in~\cite{ThesePyl}.
Common choices are the \textit{nearest neighbor PCA}~\cite{CN}
obtained by choosing the support of~$k$ as the set of the four sites 
neighboring the 
origin and the \textit{cross PCA}~\cite{CNS2}
obtained by choosing the support of~$k$ as the set made of the origin and its 
four neighboring sites (see figure~\ref{f:modelli}).

\subsection{Connection with statistical mechanics stochastic systems}
\label{s:mss}
\par\noindent
The interest of reversible PCA has already been discussed above. 
In this section we recall an interesting connection between reversible 
PCA and statistical mechanics lattice models~\cite{Derrida} and 
\cite[Section~2.4.2]{LR}. 

Consider a statistical mechanics model on the torus $\Lambda$ (periodic boundary conditions)
with configuration space $\Omega=\{-1,+1\}^\Lambda$ and Hamiltonian 
\begin{displaymath}
F(\sigma)
 =
 -\frac{1}{2}\sum_{i,j\in\Lambda}J_{ij}\sigma_i\sigma_j
 -h\sum_{i\in\Lambda}\sigma_i
\end{displaymath}
with $h\in\mathbb{R}$ and $J_{ij}$ symmetrical and translationally invariant,
that is 
$J_{ij}=J_{ji}$ and 
$J_{ij} = J_{i+s,j+s}$
for all $s\in\mathbb{Z}^2$.
The equilibrium properties of the model at inverse 
temperature $\beta$ are described by the finite--volume 
Gibbs measure 
$\nu_\beta(\sigma)=\exp\{-\beta F(\sigma)\}/\sum_{\eta\in\cc{S}}
                       \exp\{-\beta F(\eta)\}$.

The stochastic version of the model is a discrete time Markov chain 
$\sigma(0),\sigma(1),\dots,\sigma(t)$ such that its stationary measure 
is equal to the equilibrium Gibbs measure $\nu_\beta$. This can be achieved 
by choosing different transition matrices in the definition of the Markov chain. 
A very celebrated choice is the so--called 
\textit{heat--bath (Glauber)} dynamics:
at each time $t\in\mathbb{N}$
choose uniformly at random (with probability $1/|\Lambda|$) a site $i\in\Lambda$ 
and 
let $\sigma_i(t)=s$ with probability 
$f^{\textrm{HB}}_{\Theta_{i}\sigma(t-1)}(s)$
where
\begin{displaymath}
f^{\textrm{HB}}_{\sigma}(s)
=
\frac{\exp\{-\beta H(s\sigma_{\Lambda\setminus\{0\}})\}}
  {\exp\{-\beta H(s\sigma_{\Lambda\setminus\{0\}})\}
   +\exp\{-\beta H(-s\sigma_{\Lambda\setminus\{0\}})\}}
\end{displaymath}
for any $\sigma\in\Omega$, where $\sigma_{\Lambda\setminus\{0\}}$ denotes 
the restriction of the configuration $\sigma$ on $\Lambda\setminus\{0\}$.
We have that
\begin{displaymath}
\begin{array}{rcl}
{\displaystyle
 f^{\textrm{HB}}_\sigma(s)
}
 &\!\!=&\!\!
{\displaystyle
\frac{1}
  {1+\exp\{\beta[H(s\sigma_{\Lambda\setminus 0})
            -H(-s\sigma_{\Lambda\setminus 0})]\}}
=
\frac{1}
  {1+\exp\{\beta[-2s(\sum_{j\in\Lambda}J_{0j}\sigma_j+h)]\}}
\vphantom{\Bigg\}_\}}
} \\
&\!\!=&\!\!
{\displaystyle
\frac{1}{2}
\Big\{
     1+s\tanh\Big[
                  \beta
                  \Big(
                       \sum_{j\in\Lambda}J_{0j}\sigma_j+h
                  \Big)
             \Big]
\Big\}
} \\
\end{array}
\end{displaymath}
which is in the form \eqref{frev}.

Summing up, by implementing in a 
parallel fashion the heat--bath rates of a statistical mechanics 
lattice model, a reversible PCA is obtained. Notice that 
the stationary measure of the reversible PCA obtained is different 
from that of the starting statistical mechanics model. Then, 
an interesting question arises immediately: are there connections 
between the phase diagram of the starting statistical mechanics model 
and that of the resulting reversible PCA?
This question is one of the problems which is addressed in this paper. 
We recall that a similar question has been posed in~\cite{CN} 
in connection with metastability phenomena. 

\subsection{Low temperature behavior of the reversible PCA Hamiltonian}
\label{s:ham}
\par\noindent
The stationary measure $\mu_{\beta,h}$ introduced above looks like a 
finite--volume Gibbs measure with Hamiltonian
$G_{\beta,h}(\sigma)$ (see \eqref{mahG}). It is worth noting that $G_{\beta,h}$ 
cannot be thought as 
a proper statistical mechanics Hamiltonian since it depends on the temperature 
$1/\beta$. On the other hand the 
low--temperature behavior of the stationary measure of the PCA 
can be guessed by looking at the function
\begin{equation}
\label{zth}
H_h(\sigma)
=
\lim_{\beta\to\infty}G_{\beta,h}(\sigma)
=
-h\sum_{i\in\Lambda}\sigma_i
-\sum_{i\in\Lambda}
   \Big|\sum_{j\in\Lambda}k(j-i)\sigma_j+h\Big|
\end{equation}
The absolute minima of the function $H_h$ are called 
\textit{ground states} of the stationary measure for the reversible PCA.

Following~\cite{CNS1}, the difference between the 
Hamiltonian $G_{\beta,h}$ and its zero temperature limit $H_h$ can be computed. 
We have that 
\begin{equation}
\label{correzione}
G_{\beta,h}(\sigma)-H_h(\sigma)
=
-\frac{1}{\beta}
\sum_{i\in\Lambda}
 \log\Big(
          1+\exp\Big\{
                      -2\beta
                        \Big|\sum_{j\in\Lambda}k(j-i)\sigma_j+h\Big|
                \Big\}
     \Big)
+\frac{1}{\beta}|\Lambda|\log (2)
\end{equation}
for each $\beta>0$ and $\sigma\in\Omega$. Indeed, 
\begin{displaymath}
\begin{array}{rcl}
G_{\beta,h}(\sigma)-H_h(\sigma)
&\!\!\!=&\!\!\!
{\displaystyle
 -\frac{1}{\beta}\sum_{i\in\Lambda}
    \log\cosh\Big[\beta
   \Big(\sum_{j\in\Lambda}k(j-i)\sigma_j+h\Big)\Big]
 \!\!+\!\!\sum_{i\in\Lambda}\!
   \Big|\sum_{j\in\Lambda}k(j-i)\sigma_j+h\Big|
 \vphantom{\bigg\{_\bigg\}}
}
\\
&\!\!\!=&\!\!\!
{\displaystyle
 -\frac{1}{\beta}
 \sum_{i\in\Lambda}
 \Big\{
      \log\cosh\Big[\beta
      \Big(\sum_{j\in\Lambda}k(j-i)\sigma_j+h\Big)\Big]
 \vphantom{\bigg\{_\bigg\}}
}
\\
&&\!\!\!
{\displaystyle
      \phantom{aaaaaaaaaaaaaaaaa}
      -
      \log\exp\Big[\beta\Big|\sum_{j\in\Lambda}k(j-i)\sigma_j+h\Big|\Big]
 \Big\}
}
\\
\end{array}
\end{displaymath}
Hence
\begin{displaymath}
G_{\beta,h}(\sigma)-H_h(\sigma)
\!=\!
 -\frac{1}{\beta}
 \sum_{i\in\Lambda}\log 
   \frac{
         {\displaystyle
          \exp\Big[\beta\Big|\sum_{j\in\Lambda}k(j-i)\sigma_j+h\Big|\Big]
          \!+\!
          \exp\Big[-\beta\Big|\sum_{j\in\Lambda}k(j-i)\sigma_j+h\Big|\Big]
         }
        }
        {
         {\displaystyle
          2\exp\Big[\beta\Big|\sum_{j\in\Lambda}k(j-i)\sigma_j+h\Big|\Big]
         }
        } 
\end{displaymath}
which yields \eqref{correzione}.

\section{The cross PCA}
\label{s:croce}
\par\noindent
In this paper we shall study the phase diagram of the Gibbs--like 
measure associated to the \textit{cross PCA}. 
More precisely, we shall 
that $k(j)=0$ if $j$ is neither the origin
nor one of its nearest neighbors, i.e. in the cross $I$. 
In this case, since $k$ has to be symmetric 
with respect to the origin, the probability measure $f_{\sigma ;h}$ has to be 
\begin{displaymath}
 f_{\sigma ;h}(s)=\frac{1}{2}\Big\{
   1+s\tanh\Big[\beta\Big(
   k_0\sigma_0+k_1[\sigma_{e_1}+\sigma_{-e_1}]
   +k_2[\sigma_{e_2}+\sigma_{-e_2}]+h
                          \Big)\Big]\Big\}
\end{displaymath}
where $e_1$ and $e_2$ are unit vectors parallel to the coordinate 
axes of the lattice and \mbox{$k_0,k_1,k_2\in\bb{R}$}.
The constant $k_0$ is the \textit{self--interaction intensity}.
To sum up, the \emph{cross PCA} is a family of PCA dynamics parameterized by $k_0,k_1,k_2,\beta,h$.

Note that for the cross model the Hamiltonian $G_{\beta,h}$ 
defining the stationary 
Gibbs--like measure is given by 
\begin{equation}
\label{ham-cross}
\begin{array}{rl}
G_{\beta,h}(\sigma)
=
{\displaystyle
 -h\sum_{i\in\Lambda}\sigma_i
 -\frac{1}{\beta}\sum_{i\in\Lambda}
    \log\cosh\Big[\beta
   \Big(
}
&
{\displaystyle
   \!\!\!
   k_0\sigma_i+k_1[\sigma_{i+e_1}+\sigma_{i-e_1}]
}
\\
&
{\displaystyle
   \!\!\!
 +k_2[\sigma_{i+e_2}+\sigma_{i-e_2}]+h
       \Big)\Big]
}
\end{array}
\end{equation}
The Hamiltonian can be rewritten as 
\begin{equation}
\label{ham-cross010}
G_{\beta,h}(\sigma)
=
\sum_{i\in\Lambda}
G_{\beta,h,i}(\sigma)
\end{equation}
where
\begin{equation}
\label{ham-cross015}
G_{\beta,h,i}(\sigma)
=
G_{\beta,h}^0(\Theta_i\sigma)
\end{equation}
and 
\begin{equation}
\label{ham-cross020}
\begin{array}{rcl}
G_{\beta,h}^0(\sigma)
&\!\!\!=&\!\!\!
{\displaystyle
 -\frac{1}{5}h
 [
  \sigma_0+\sigma_{e_1}+\sigma_{-e_1}+\sigma_{e_2}+\sigma_{-e_2}
 ]
 \vphantom{\bigg\{_\big\}}
}
\\
&&\!\!\!
{\displaystyle
 -\frac{1}{\beta}
    \log\cosh\{\beta
   (
    k_0\sigma_0+k_1[\sigma_{e_1}+\sigma_{-e_1}]
    +k_2[\sigma_{e_2}+\sigma_{-e_2}]+h
   )\}
}
\end{array}
\end{equation}
Note that, 
for any $i\in\Lambda$, $G_{\beta,h,i}$ 
is the contribution of the cross centered at the site $i$ of the torus 
$\Lambda$ to the Hamiltonian of the system. 

\begin{figure}
\setlength{\unitlength}{0.7pt}
\begin{picture}(200,100)(-190,-10)  
\thinlines
\multiput(-20,-5)(20,0){18}{\line(0,1){100}}
\multiput(-30,5)(0,20){5}{\line(1,0){360}}
\thicklines
\linethickness{0.7mm}
\put(-20,85){\circle*{10}}
\put(20,85){\circle*{10}}
\put(40,85){\circle*{10}}
\qbezier(20,85)(30,85)(40,85)
\put(80,65){\circle*{10}}
\put(100,85){\circle*{10}}
\qbezier(80,65)(90,75)(100,85)
\put(140,85){\circle*{10}}
\put(180,85){\circle*{10}}
\qbezier(140,85)(160,85)(180,85)
\put(220,65){\circle*{10}}
\put(240,85){\circle*{10}}
\put(260,65){\circle*{10}}
\qbezier(220,65)(230,75)(240,85)
\qbezier(220,65)(240,65)(260,65)
\qbezier(240,85)(250,75)(260,65)
\put(-20,45){\circle*{10}}
\put(-20,25){\circle*{10}}
\put(0,25){\circle*{10}}
\qbezier(-20,45)(-20,35)(-20,25)
\qbezier(-20,25)(-10,25)(0,25)
\put(40,25){\circle*{10}}
\put(60,25){\circle*{10}}
\put(80,25){\circle*{10}}
\qbezier(40,25)(60,25)(80,25)
\put(120,25){\circle*{10}}
\put(140,25){\circle*{10}}
\put(160,25){\circle*{10}}
\put(140,45){\circle*{10}}
\qbezier(120,25)(140,25)(160,25)
\qbezier(140,25)(140,35)(140,45)
\put(200,25){\circle*{10}}
\put(220,45){\circle*{10}}
\put(220,5){\circle*{10}}
\put(240,25){\circle*{10}}
\qbezier(200,25)(210,35)(220,45)
\qbezier(200,25)(210,15)(220,5)
\qbezier(220,5)(230,15)(240,25)
\qbezier(220,45)(230,35)(240,25)
\put(280,25){\circle*{10}}
\put(300,45){\circle*{10}}
\put(300,25){\circle*{10}}
\put(300,5){\circle*{10}}
\put(320,25){\circle*{10}}
\qbezier(280,25)(300,25)(320,25)
\qbezier(300,5)(300,25)(300,45)
\end{picture}
\caption{Schematic representation of the coupling constants: from the left to
the right and from the top to the bottom the couplings 
$J_{.}$, 
$J_{\langle\rangle_1}$, 
$J_{\langle\langle\rangle\rangle}$,
$J_{\langle\langle\langle \rangle\rangle\rangle_1}$, 
$J_{\triangle_1}$,
$J_{\llcorner}$,
$J_{\AC_1}$,
$J_{\perp_1}$,
$J_{\diamondsuit}$,
and
$J_+$
are depicted.}
\label{f:accoppiamenti}
\end{figure}

\subsection{Coupling constants for the Hamiltonian of the cross PCA}
\label{s:ham-croce}
\par\noindent
In statistical mechanics systems the Hamiltonian is usually written as 
a sum of potentials each of them being the product of the spin variables 
over some subset of the lattice multiplied by a constant called 
\textit{coupling constant}. 
For any $\sigma\in\Omega$ and any $I\subset\Lambda$ we set 
\begin{equation}
\label{pot001}
\sigma_I
=
\prod_{i\in I}\sigma_i
\end{equation}
In this section we prove that the 
Hamiltonian \eqref{ham-cross} for the cross PCA can be written in the form 
\begin{equation}
\label{pot005}
\begin{array}{rcl} 
{\displaystyle 
 G_{\beta,h}(\sigma)
} 
&\!\!=&\!\!
{\displaystyle{ 
 -J_{.} 
 \sum_{i\in\Lambda}\sigma_i 
 -J_{\langle\rangle_1} 
 \sum_{\langle\rangle_1}\sigma_{\langle\rangle_1}
 -J_{\langle\rangle_2} 
 \sum_{\langle\rangle_2}\sigma_{\langle\rangle_2} 
 -J_{\langle\langle\rangle\rangle} 
 \sum_{\langle\langle\rangle\rangle}\sigma_{\langle\langle\rangle\rangle}
 -J_{\langle\langle\langle \rangle\rangle\rangle_1} 
 \sum_{\langle\langle\langle\rangle\rangle\rangle_1}
          \sigma_{\langle\langle\langle\rangle\rangle\rangle_1}
 \vphantom{\bigg\{_\bigg\}}
}}\\ 
&\!\!&\!\!
{\displaystyle{ 
 -J_{\langle\langle\langle \rangle\rangle\rangle_2} 
 \sum_{\langle\langle\langle\rangle\rangle\rangle_2}
          \sigma_{\langle\langle\langle\rangle\rangle\rangle_2}
 -J_{\triangle_1} 
 \sum_{\triangle_1}\sigma_{\triangle_1}
 -J_{\triangle_2} 
 \sum_{\triangle_2}\sigma_{\triangle_2}
 -J_{\llcorner} 
 \sum_{\llcorner}\sigma_\llcorner
 -J_{\AC_1} 
 \sum_{\AC_1}\sigma_{\AC_1}
 \vphantom{\bigg\{_\bigg\}}
}}\\ 
&\!\!&\!\!
{\displaystyle
 -J_{\AC_2} 
 \sum_{\AC_2}\sigma_{\AC_2}
 -J_{\perp_1} 
 \sum_{\perp_1}\sigma_{\perp_1}
 -J_{\perp_2} 
 \sum_{\perp_2}\sigma_{\perp_2}
 -J_{\diamondsuit} 
 \sum_{\diamondsuit}\sigma_\diamondsuit
 -J_{+} 
 \sum_{+}\sigma_+
} 
\\ 
\end{array} 
\end{equation}
where the meaning of each term is illustrated in figure~\ref{f:accoppiamenti}
and its expression as
function of~$h$, $\beta$, $k_0$, $k_1$, and~$k_2$
is given in the Appendix~\ref{s:accoppiamenti}. Note that~$1$ and~$2$ mean, 
respectively, oriented along the $1$ and the $2$ coordinate direction 
(namely, horizontal and vertical).

We calculate the coefficients $J$'s by using 
\cite[equations~(6) and (7)]{CS96} 
(see also \cite[equations~(3.1) and (3.2)]{HK} and \cite{CNP}). 
More precisely, given $f:\{-1,+1\}^V\to\bb{R}$, with $V\subset\bb{Z}^2$ 
finite, we have that for any $\sigma\in\{-1,+1\}^V$
\begin{equation}
\label{pot000}
f(\sigma)
=
\sum_{I\subset V} C_I\prod_{i\in I}\sigma_i
\end{equation}
with the coefficients $C_I$'s given by 
\begin{equation}
\label{pot010}
C_I
=
\frac{1}{2^{|V|}}
\sum_{\sigma\in\{-1,+1\}^V}f(\sigma)\prod_{i\in I}\sigma_i
\end{equation}

\begin{figure}[t]
\begin{picture}(100,120)(-100,0)
\includegraphics[height=5.cm,angle=0]{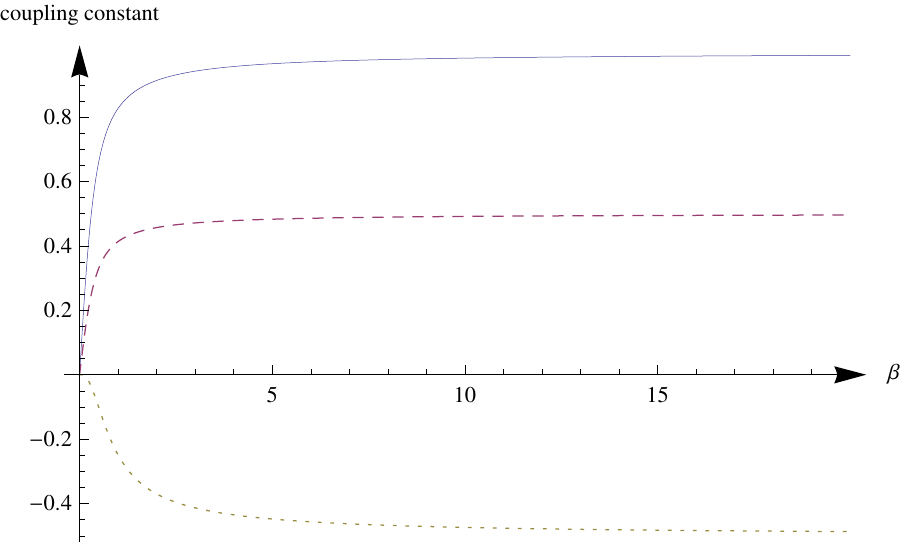}
\end{picture}
\caption{The coupling constants 
$J_{\langle\langle\rangle\rangle}$ (solid), 
$J_{\langle\langle\langle\rangle\rangle\rangle_1}=
 J_{\langle\langle\langle\rangle\rangle\rangle_2}$ (dashed), and
$J_{\diamond}$ (dotted).
are plotted
at $h=0$, $k_0=0$, and $k_1=k_2=1$ as a function of $\beta$.}
\label{f:acc-beta}
\end{figure}

By expanding the function $-G_{\beta,h}^0$ defined in the 
cross centered at the origin we obtain, by exploiting the symmetry of the 
cross, the seventeen different coefficients 
$J^0_{\bullet}$, 
$J^0_{\circ_1}$, 
$J^0_{\circ_2}$, 
$J^0_{\langle\rangle_1}$, 
$J^0_{\langle\rangle_2}$, 
$J^0_{\langle\langle\rangle\rangle}$,
$J^0_{\langle\langle\langle \rangle\rangle\rangle_1}$, 
$J^0_{\langle\langle\langle \rangle\rangle\rangle_2}$, 
$J^0_{\triangle_1}$,
$J^0_{\triangle_2}$,
$J^0_{\llcorner}$,
$J^0_{\AC_1}$,
$J^0_{\AC_2}$,
$J^0_{\perp_1}$,
$J^0_{\perp_2}$,
$J^0_{\diamondsuit}$,
and
$J^0_+$
whose meaning is the same as for the $J$'s introduced above
(see also figure~\ref{f:accoppiamenti}) but for the single site 
coefficients 
$J^0_{\bullet}$, $J^0_{\circ_1}$, 
and $J^0_{\circ_2}$ which refer, respectively, 
to the center and to the peripheral sites of the cross on the~$1$ and 
on the~$2$ direction. 
The coefficients we get are listed in 
the Appendix~\ref{s:accoppiamenti}. 

By exploiting, now, the translational invariance of the cross PCA 
Hamiltonian $G_{\beta,h}$ we have that 
\begin{equation}
\label{pot020}
\begin{array}{lllll}
J_{.}=J^0_{\bullet}+2J^0_{\circ_1}+2J^0_{\circ_2},
&
J_{\langle\rangle_1}=2J^0_{\langle\rangle_1},
&
J_{\langle\rangle_2}=2J^0_{\langle\rangle_2}, 
&
J_{\langle\langle\rangle\rangle}=2J^0_{\langle\langle\rangle\rangle},
&
J_{\langle\langle\langle \rangle\rangle\rangle_1}=
J^0_{\langle\langle\langle \rangle\rangle\rangle_1}, 
\\
J_{\langle\langle\langle \rangle\rangle\rangle_2}=
J^0_{\langle\langle\langle \rangle\rangle\rangle_2}, 
&
J_{\triangle_1}=J^0_{\triangle_1},
&
J_{\triangle_2}=J^0_{\triangle_2},
&
J_{\llcorner}=J^0_{\llcorner},
&
J_{\AC_1}=J^0_{\AC_1},
\\
J_{\AC_2}=J^0_{\AC_2},
&
J_{\perp_1}=J^0_{\perp_1},
&
J_{\perp_2}=J^0_{\perp_2},
&
J_{\diamondsuit}=J^0_{\diamondsuit},
&
J_{+}=J^0_{+}
\end{array}
\end{equation}

\begin{figure}[t]
\begin{picture}(100,120)(0,0)
\put(0,0)
{
 \includegraphics[height=5cm,angle=0]{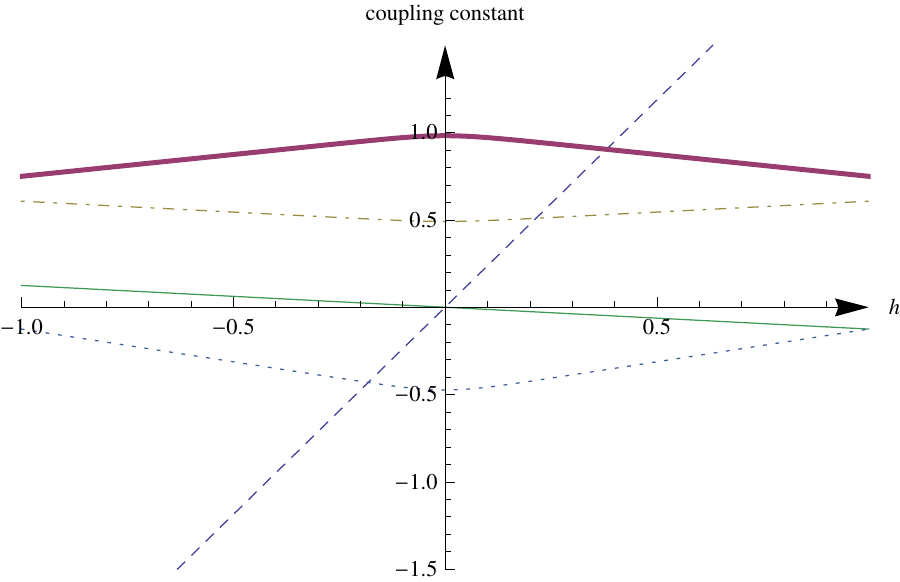}
}
\put(240,0)
{
 \includegraphics[height=5cm,angle=0]{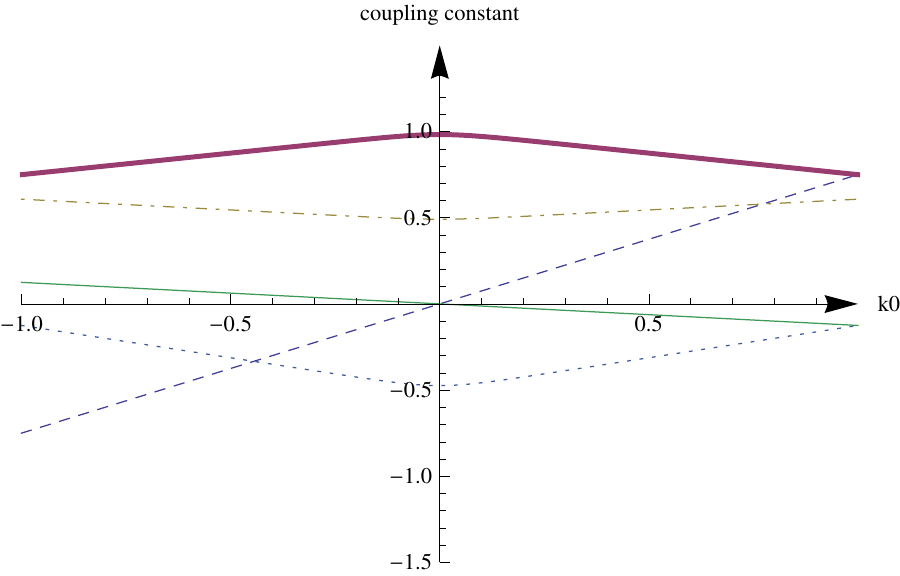}
}
\end{picture}
\caption{Left: the coupling constants 
$J_{\cdot}$ (dashed), 
$J_{\langle\langle\rangle\rangle}$ (solid thick), 
$J_{\langle\langle\langle\rangle\rangle\rangle_1}=
 J_{\langle\langle\langle\rangle\rangle\rangle_2}$ (dot--dashed), 
$J_{\triangle_1}=J_{\triangle_2}$ (solid thin), 
and
$J_{\diamond}$ (dotted)
are plotted
at $\beta=10$, $k_0=0$, and $k_1=k_2=1$ as a function of $h$.
Right:
the coupling constants 
$J_{\langle\rangle_1}=J_{\langle\rangle_2}$ (dashed), 
$J_{\langle\langle\rangle\rangle}$ (solid thick), 
$J_{\langle\langle\langle\rangle\rangle\rangle_1}=
 J_{\langle\langle\langle\rangle\rangle\rangle_2}$ (dot--dashed), 
$J_{\perp_1}=J_{\perp_2}$ (solid thin), 
and
$J_{\diamond}$ (dotted)
are plotted
at $\beta=10$, $h=0$, and $k_1=k_2=1$ as a function of $k_0$.}
\label{f:acc-accak0}
\end{figure}

The coupling constants $J$'s depend on the parameters of the model, 
namely, $\beta$, $h$, $k_0$, $k_1$, and $k_2$, see also the 
explicit expressions given in Appendix~\ref{s:accoppiamenti}. To give 
an idea of their typical values we plot some of them in the
figures~\ref{f:acc-beta} and \ref{f:acc-accak0}.

In figure~\ref{f:acc-beta} we plot the coupling constants 
at $h=0$, $k_0=0$, and $k_1=k_2=1$ as a function of $\beta$.
Note that with such a choice of the parameters the sole non zero 
couplings are 
$J_{\langle\langle\rangle\rangle}$, 
$J_{\langle\langle\langle\rangle\rangle\rangle_1}$, 
$J_{\langle\langle\langle\rangle\rangle\rangle_2}$, and
$J_{\diamond}$.
Moreover, since $k_1=k_2$, it turns out that 
$J_{\langle\langle\langle\rangle\rangle\rangle_1}=
 J_{\langle\langle\langle\rangle\rangle\rangle_2}$.

In figure~\ref{f:acc-accak0} on the left we plot the coupling constants 
at $\beta=10$, $k_0=0$, and $k_1=k_2=1$ as a function of $h$.
Note that, from the graphs in figure~\ref{f:acc-beta}, it results 
that the coupling constants are approximatively constant for $\beta\ge5$. 
Note that with such a choice of the parameters the sole non zero 
couplings are 
$J_{\cdot}$, 
$J_{\langle\langle\rangle\rangle}$, 
$J_{\langle\langle\langle\rangle\rangle\rangle_1}$, 
$J_{\langle\langle\langle\rangle\rangle\rangle_2}$, 
$J_{\triangle_1}$, 
$J_{\triangle_2}$, 
and
$J_{\diamond}$.
Moreover, since $k_1=k_2$, it turns out that 
$J_{\langle\langle\langle\rangle\rangle\rangle_1}=
 J_{\langle\langle\langle\rangle\rangle\rangle_2}$
and 
$J_{\triangle_1}=J_{\triangle_2}$.

In figure~\ref{f:acc-accak0} on the right we plot the coupling constants 
at $\beta=10$, $h=0$, and $k_1=k_2=1$ as a function of $k_0$.
Note that with such a choice of the parameters the sole non zero 
couplings are 
$J_{\langle\rangle_1}$, 
$J_{\langle\rangle_2}$, 
$J_{\langle\langle\rangle\rangle}$, 
$J_{\langle\langle\langle\rangle\rangle\rangle_1}$, 
$J_{\langle\langle\langle\rangle\rangle\rangle_2}$, 
$J_{\perp_1}$, 
$J_{\perp_2}$, 
and
$J_{\diamond}$.
Moreover, since $k_1=k_2$, it turns out that 
$J_{\langle\rangle_1}=J_{\langle\rangle_2}$, 
$J_{\langle\langle\langle\rangle\rangle\rangle_1}=
 J_{\langle\langle\langle\rangle\rangle\rangle_2}$
and 
$J_{\perp_1}=J_{\perp_2}$.

\subsection{Ground states of the cross PCA}
\label{s:fond}
\par\noindent
Assume that the side length 
of the torus $\Lambda$ is an even number. Furthermore, we assume that $|k_0| < 2(k_1+k_2)$ in order for the self-interaction not to be able to change the majority of the cross configuration.  
Recalling the definition of ground states given in Section~\ref{s:ham}, we have
that the ground states of the cross PCA are the absolute minima of the 
function 
\begin{equation}
\label{zth-cross}
\begin{array}{rl}
H_h(\sigma)
=
{\displaystyle
 -h\sum_{i\in\Lambda}\sigma_i
 -\sum_{i\in\Lambda}
   \Big|
}
&
{\displaystyle
   \!\!\!
   k_0\sigma_i+k_1[\sigma_{i+e_1}+\sigma_{i-e_1}]
}
\\
&
{\displaystyle
 \phantom{aaaa}
 +k_2[\sigma_{i+e_2}+\sigma_{i-e_2}]+h
       \Big|
}
\end{array}
\end{equation}
which can be rewritten as 
\begin{equation}
\label{zth-cross-015}
H_h(\sigma)
=
 \sum_{i\in\Lambda}
 H_{h,i}(\sigma)
\end{equation}
where 
\begin{equation}
\label{zth-cross-010}
\begin{array}{rl}
\!\!\!
\!\!\!
H_{h,i}(\sigma)
=&\!\!\!\!
{\displaystyle
 -
  \Big[  
       \frac{1}{5}h[\sigma_i+\sigma_{i+e_1}+\sigma_{i-e_1}
       +\sigma_{i+e_2}+\sigma_{i-e_2}]
}
\\
&
\phantom{aa}
   +|k_0\sigma_i+k_1[\sigma_{i+e_1}+\sigma_{i-e_1}]
     +k_2[\sigma_{i+e_2}+\sigma_{i-e_2}]+h|
  \Big]
\end{array}
\end{equation}
Remark that 
\begin{displaymath}
\lim_{\beta\to\infty}G_{\beta,h,i}(\sigma)=H_{h,i}(\sigma)
\end{displaymath}
for each $h\in\bb{R}$, $i\in\Lambda$, and $\sigma\in\Omega$.
We also note that 
\begin{equation}
\label{zth-cross-020}
H_h(\sigma)=H_{-h}(-\sigma)
\end{equation}
for any $h\in\bb{R}$ and $\sigma\in\Omega$, where $-\sigma$ denotes 
the configuration obtained by flipping the sign of all the spins of $\sigma$.
By \eqref{zth-cross-020} we can bound our discussion to the case $h\ge0$ and
deduce a posteriori the structure of the ground states for $h<0$. 

Moreover, we notice that, since $H_h$ is in the form 
\eqref{zth-cross-015}, a global minimum of $H_h$ is realized by minimizing each $H_{h,i}(\sigma)$ 
for any $i\in\Lambda$.  

\begin{figure}
\setlength{\unitlength}{0.7pt}
\begin{picture}(200,240)(-220,-10)  
\qbezier[30](0,100)(100,100)(200,100)
\qbezier[30](100,0)(100,100)(100,200)
\put(200,100){\vector(1,0){5}}
\put(100,200){\vector(0,1){5}}
\thicklines
\put(100,100){\line(1,0){100}}
\put(100,100){\line(-1,1){90}}
\put(100,100){\line(-1,-1){90}}

\put(200,85){${k_0}$}
\put(105,200){${h}$}
\put(140,150){${\puno}$}
\put(140,50){${\muno}$}
\put(30,96){${\chess_\textrm{e},\,\chess_\textrm{o}}$}
\end{picture}
\caption{Expected 
zero temperature phase diagram of the stationary measure of the cross PCA.
On the thick lines the ground states of the adjacent
regions coexist. At the origin the four ground states coexist.}
\label{f:ground}
\end{figure}
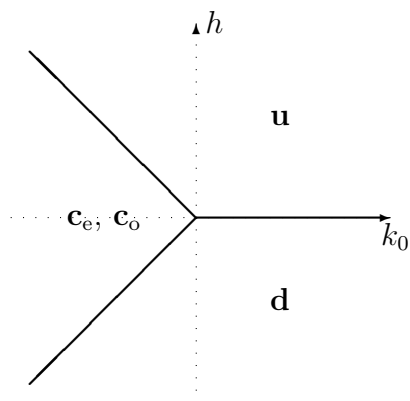

We discuss the structure of the ground states of the cross PCA under the 
assumption $k_1,k_2>0$ and consider the following cases (see 
figure~\ref{f:ground}). 

\par\noindent\textit{Case $h>0$ and $k_0\ge0$.\/}
The minimum of $H_{h,i}$ is attained at the cross configuration having 
all the spins equal to plus one. Hence the unique absolute minimum 
of $H_h$ is the configuration $\puno$ such that 
$\puno(i)=+1$ for all $i\in\Lambda$. 

\par\noindent\textit{Case $h>-k_0$ and $k_0<0$.\/}
If the spin at $i$ is plus, since $h+k_0>0$, $H_{h,i}$ is then minimal 
provided
all the other spins in the cross are equal to $+1$
and it is equal to $-h-|k_0+2(k_1+k_2)+h|$. 
If the spin at $i$ is minus, since $h-k_0>0$, $H_{h,i}$ is then minimal 
provided all the other spins in the cross are equal to $+1$
and it is equal to 
$-3h/5-|-k_0+2(k_1+k_2)+h|$. 
Is not difficult to prove that  
\begin{displaymath}
-h-|k_0+2(k_1+k_2)+h|\le
-3h/5-|-k_0+2(k_1+k_2)+h|
\textrm{ if and only if }
h\ge -5k_0
\end{displaymath}
We can than conclude that in the region $k_0<0$ and $h>-5k_0$ 
the ground state of the cross PCA is the configuration $\puno$.

In the region $k_0<0$ and $0<h<-5k_0$ the situation is more 
complicated and we expect that the line $h=-k_0$ is the 
boudary between the regions where the ground states are 
respectively $\puno$ and 
the pair 
$\chess_\textrm{e}$ and $\chess_\textrm{o}$ with 
$\chess_\textrm{e}$ the checkerboard configuration with pluses on the 
even sub--lattice of $\Lambda$ and minuses on its complement, while 
$\chess_\textrm{o}$ is the corresponding spin--flipped configuration.
Indeed, 
we can write $H_h(\puno)=-|\Lambda|[h+|k_0+2k_1+2k_2+h|]$
and, hence, 
\begin{displaymath}
H_h(\puno)=
\left\{
\begin{array}{ll}
-|\Lambda|[k_0+2(k_1+k_2)+2h]
&
\;\;\textrm{ if }
k_0+2k_1+2k_2+h>0
\\
-|\Lambda|[-k_0-2(k_1+k_2)]
&
\;\;\textrm{ if }
k_0+2k_1+2k_2+h<0
\end{array}
\right.
\end{displaymath}
Moreover, recalling $k_0<0$,
$H_h(\chess_\textrm{e})=H_h(\chess_\textrm{o})
=
-(|\Lambda|/2)[|k_0-2k_1-2k_2+h|
                +|-k_0+2k_1+2k_2+h|]
=
-(|\Lambda|/2)[|k_0-2k_1-2k_2+h|
                +(-k_0+2k_1+2k_2+h)]
$.
Hence, 
\begin{displaymath}
H_h(\chess_\textrm{e})=H_h(\chess_\textrm{o})=
\left\{
\begin{array}{ll}
-|\Lambda|h
&
\;\;\textrm{ if }
k_0-2k_1-2k_2+h>0
\\
-|\Lambda|[-k_0+2k_1+2k_2]
&
\;\;\textrm{ if }
k_0-2k_1-2k_2+h<0
\end{array}
\right.
\end{displaymath}
From these explicit expressions, by discussing separately the three 
cases
$h>-k_0+2(k_1+k_2)$, 
$-k_0+2(k_1+k_2)>h>-k_0-2(k_1+k_2)$, 
and
$-k_0-2(k_1+k_2)>h$, 
it follows that 
$H_h(\chess_\textrm{e})=H_h(\chess_\textrm{o})<H_h(\puno)$ 
if and only if $h+k_0<0$.

We now discuss those cases in which the external field $h$ is equal 
to zero. We first note that in this case the term $H_{h,i}$ reduces to 
\begin{equation}
\label{zth-cross-030}
H_{0,i}(\sigma)
=-|k_0\sigma_i+k_1[\sigma_{i+e_1}+\sigma_{i-e_1}]
     +k_2[\sigma_{i+e_2}+\sigma_{i-e_2}]|
\end{equation}

\par\noindent\textit{Case $h=0$ and $k_0>0$.\/}
The minimum of $H_{0,i}$ is attained at the cross configuration having 
all the spins equal to plus one or all equal to minus one. 
Hence the set of ground states is made of the two configurations 
$\puno$ and $\muno$ with this last one such that 
$\muno(i)=-1$ for all $i\in\Lambda$. 

\par\noindent\textit{Case $h=0$ and $k_0=0$.\/}
The minimum of $H_{0,i}$ is attained at the cross configuration having 
all the spins equal to plus one or all equal to minus one on the neighbors 
of the center and with the spin at the center which can be, in any case, 
either plus or minus. 
Hence the set of ground states is made of the four configurations 
$\puno$, $\muno$, $\chess_\textrm{e}$, and $\chess_\textrm{o}$.

\par\noindent\textit{Case $h=0$ and $k_0<0$.\/}
The minimum of $H_{0,i}$ is attained at the cross configuration having 
the spin at the center equal to plus one and the others equal to minus one 
and at the spin--flipped cross configuration.
Hence the set of ground states is made of the two configurations 
$\chess_\textrm{e}$ and $\chess_\textrm{o}$.

\par\noindent\textit{Case $h<0$.\/} The set of ground states 
can be easily discussed as for $h>0$ by using the 
property \eqref{zth-cross-020}.

\subsection{Phase diagram}
\label{s:phase}
\par\noindent
The goal of this paper is the study of the effect of the 
self--interaction parameter~$k_0$ on the phase diagram of the Gibbs--like 
measure \eqref{gibbs} with Hamiltonian \eqref{ham-cross} associated with 
the cross PCA. 
In other words we want to study how the ground states structure 
depicted in figure~\ref{f:ground} is deformed when a small 
non--zero temperature is considered.
One interesting feature 
is the connection with the well know 
Blume--Capel model which has been widely discussed in the introduction, 
see Section~\ref{s:syn}.

As recalled in Section~\ref{s:syn} such a phase diagram is 
strictly connected with the structure of the stationary 
states of the PCA. On this subject 
very few rigorous results have been established:
\begin{itemize}
\item[--]
it has been remarked  in~\cite{KV} and  in~\cite[Remark~4.1]{DpLR} 
and proven more 
accurately~\cite[Th.~4.2.5 Section 4.2.2, Proposition~4.3.4 Section 4.3.3]{ThesePyl} 
 that at~$h=k_0=0$ the model can be 
decoupled into two independent systems defined on the even and on the odd 
part of the lattice and that the phase diagram is the same as 
that of two independent nearest--neighbor Ising models. The set of 
Gibbs distributions consists in the convex hull generated
by the four Gibbs distributions respectively close to 
$\puno$, $\muno$, $\chess_\textrm{e}$, and $\chess_\textrm{o}$ and obtained by taking those as boundary conditions and 
considering the thermodynamic limit. 
\item[--]
For $h=0$, $k_1\neq 0$, $k_2\neq 0$ and for any $k_0$, it has 
been proven in~\cite[Proposition~4.1]{DpLR} that for $T$
smaller than a critical value, a phase transition for the 
Gibbs--like measure associated with the PCA occurs. 
\end{itemize}

Consider, now, the three--dimensional space $k_0$--$h$--$T$, where we 
recall $T=1/\beta$ 
is the temperature. The rigorous results quoted above are related 
to some subparts of such a space. What we want to do is to study the phase 
diagram in the whole space and precise which phases are at play.
Obtaining rigorous results of this kind is an absolutely difficult task.
In this paper we shall give an idea of the phase diagram in this space 
via Mean Field computations, see Section~\ref{s:mf}.


\section{Mean Field phase diagram}
\label{s:mfpd}
\par\noindent
In this section we shall study the phase diagram 
of the Gibbs--like measure associated with 
the cross PCA 
via a suitable Mean Field approximation.

\subsection{The Mean Field free energy}
\label{s:mf}
\par\noindent
The Mean Field approximation relies on considering non--correlated 
degrees of freedom. In other words the equilibrium density matrix 
$\varrho_\Lambda$
is supposed to be factorized as the product of one--site transition matrices
$\varrho_x$ for each $x\in\Lambda$. 
Moreover, inspired by the structure of the ground states discussed 
above, we partition the squared torus $\Lambda$ in its even and odd 
components $\Lambda_\rr{e}$ and $\Lambda_\rr{o}$, respectively, and 
assume that with each site of the even component is associated 
the single site matrix $\varrho_\rr{e}$ 
and 
with each site of the odd component is associated 
the single site matrix $\varrho_\rr{o}$. 
In other words we assume $\varrho_x=\varrho_\rr{e}$ for $x\in\Lambda_\rr{e}$
and 
$\varrho_x=\varrho_\rr{o}$ for $x\in\Lambda_\rr{o}$, that is to say 
we assume 
the density matrix to be in the form 
\begin{equation}
\label{mf000}
\varrho_\Lambda(\sigma)
=
\prod_{x\in\Lambda_\rr{e}}\varrho_\rr{e}(\sigma_x)
\prod_{y\in\Lambda_\rr{o}}\varrho_\rr{o}(\sigma_y)
\end{equation}
for any configuration $\sigma\in\Omega$. 
The next step is that of 
computing, under such assumption, the free energy \cite{CCGM}
\begin{equation}
\label{mf010}
f=\frac{1}{|\Lambda|}
  \Big(U-\frac{1}{\beta}S\Big)
\end{equation}
where
\begin{equation}
\label{mf020}
U:=\sum_{\sigma\in\Omega}\varrho_\Lambda(\sigma) G_{\beta,h}(\sigma)
\;\;\;\textrm{ and }\;\;\;
S:=-\sum_{\sigma\in\Omega}\varrho_\Lambda(\sigma)\log\varrho_\Lambda(\sigma)
\end{equation}
are, respectively, the \textit{internal energy} and the \textit{entropy}.

We now compute the entropy 
under the Mean Field hypothesis. First of all we note that 
\begin{displaymath}
-S
=
\sum_{\sigma\in\Omega}
\prod_{x\in\Lambda_\rr{e}}\varrho_\rr{e}(\sigma_x)
\prod_{y\in\Lambda_\rr{o}}\varrho_\rr{o}(\sigma_y)
\log\Big(
         \prod_{w\in\Lambda_\rr{e}}\varrho_\rr{e}(\sigma_w)
         \prod_{z\in\Lambda_\rr{o}}\varrho_\rr{o}(\sigma_z)
    \Big)
\end{displaymath}
and, hence, 
\begin{displaymath}
-S
=
\sum_{w\in\Lambda_\rr{e}}
\sum_{\sigma\in\Omega}
\prod_{x\in\Lambda_\rr{e}}\varrho_\rr{e}(\sigma_x)
\prod_{y\in\Lambda_\rr{o}}\varrho_\rr{o}(\sigma_y)
\log\varrho_\rr{e}(\sigma_w)
+
\sum_{z\in\Lambda_\rr{o}}
\sum_{\sigma\in\Omega}
\prod_{x\in\Lambda_\rr{e}}\varrho_\rr{e}(\sigma_x)
\prod_{y\in\Lambda_\rr{o}}\varrho_\rr{o}(\sigma_y)
\log\varrho_\rr{o}(\sigma_z)
\end{displaymath}
Finally, we get
\begin{equation}
\label{mf040}
-S
=
\frac{|\Lambda|}{2}
\Big(
     \sum_{s=\pm1}\varrho_\rr{e}(s)\log\varrho_e(s)
     +
     \sum_{s=\pm1}\varrho_\rr{o}(s)\log\varrho_o(s)
\Big)
\end{equation}
We define now the \textit{even} and the \textit{odd site} 
magnetizations 
\begin{equation}
\label{mf050}
m_\rr{e}=\sum_{s=\pm1}s\varrho_\rr{e}(s)
\;\;\;\textrm{ and }\;\;\;
m_\rr{o}=\sum_{s=\pm1}s\varrho_\rr{o}(s)
\end{equation}
Note that, recalling $\varrho_\rr{r}(+1)+\varrho_\rr{r}(-1)=1$ for 
$\rr{r}=\rr{e},\rr{o}$, we have that  
\begin{equation}
\label{mf060}
\varrho_\rr{r}(+1)=\frac{1+m_\rr{r}}{2}
\;\;\;\textrm{ and }\;\;\;
\varrho_\rr{r}(-1)=\frac{1-m_\rr{r}}{2}
\end{equation}
with $\rr{r}=\rr{e},\rr{o}$.
By \eqref{mf040} and \eqref{mf060} we get, for the Mean Field entropy, 
the expression 
\begin{equation}
\label{mf070}
\begin{array}{rcl}
-S
&\!\!=&\!\!
{\displaystyle
 \frac{|\Lambda|}{2}
 \Big\{
 \frac{1+m_\rr{e}}{2}\log\frac{1+m_\rr{e}}{2}
 +\frac{1-m_\rr{e}}{2}\log\frac{1-m_\rr{e}}{2}
 \vphantom{\bigg\{_\}}
}
\\
&\!\!&\!\!
{\displaystyle
 \phantom{
          \frac{|\Lambda|}{2}
          \Big\{
         }
 +\frac{1+m_\rr{o}}{2}\log\frac{1+m_\rr{o}}{2}
 +\frac{1-m_\rr{o}}{2}\log\frac{1-m_\rr{o}}{2}
 \Big\}
}
\end{array}
\end{equation}

The computation of the internal energy in the Mean Field approximation 
is more delicate and the expansion for the Hamiltonian discussed in 
Section~\ref{s:ham-croce} will be used.
First of all recall \eqref{mf020} and note that 
\begin{displaymath}
U
=
\sum_{x\in\Lambda}
\sum_{\sigma\in\Omega}
  \varrho_\Lambda(\sigma) G_{\beta,h,x}(\sigma)
=
\sum_{x\in\Lambda}
\sum_{\newatop{\sigma_y=\pm1:}
              {y\in C(x)}}
   \Big(\prod_{y\in C(x)}\varrho_y(\sigma_y)\Big) G_{\beta,h,x}(\sigma)
\end{displaymath}
where we have used \eqref{ham-cross010}, set
$C(x):=\{x,x+e_1,x-e_1,x+e_2,x-e_2\}$
for each $x\in\Lambda$, and recalled that $G_{\beta,h,x}(\sigma)$ 
depends only on the spins $\sigma_y$ with $y\in C(x)$. 

Now, since 
$0=(0,0)\in\Lambda_\rr{e}$ and
$(1,0)\in\Lambda_\rr{o}$, by exploiting the translation invariance 
of the Hamiltonian and the structure of the Mean Field transition 
matrices, we have that 
\begin{displaymath}
U
=
\frac{|\Lambda|}{2}
\Big[
\sum_{\newatop{\sigma_y=\pm1:}
              {y\in C(0)}}
   \Big(\prod_{y\in C(0)}\varrho_y(\sigma_y)\Big) G_{\beta,h,0}(\sigma)
+
\sum_{\newatop{\sigma_y=\pm1:}
              {y\in C((1,0))}}
   \Big(\prod_{y\in C((1,0))}\varrho_y(\sigma_y)\Big) G_{\beta,h,(1,0)}(\sigma)
\Big]
\end{displaymath}
We finally get 
\begin{equation}
\label{mf080}
U
=
\frac{|\Lambda|}{2}(u_\rr{e}+u_\rr{o})
\end{equation}
with 
\begin{displaymath}
u_\rr{e}=
\sum_{\newatop{\sigma_y=\pm1:}
              {y\in C(0)}}
   \Big(\prod_{y\in C(0)}\varrho_y(\sigma_y)\Big) G_{\beta,h,0}(\sigma)
\;\;\;\textrm{ and }\;\;\;
u_\rr{o}=
\sum_{\newatop{\sigma_y=\pm1:}
              {y\in C((1,0))}}
   \Big(\prod_{y\in C((1,0))}\varrho_y(\sigma_y)\Big) G_{\beta,h,(1,0)}(\sigma)
\end{displaymath}

The coefficients $J^0$'s introduced below \eqref{pot010} and 
listed in the Appendix~\ref{s:accoppiamenti}, have 
been obtained by expanding the function $G^0_{\beta,h}$. This remark 
and a straightforward computation yield to the equation 
\begin{equation}
\label{mf090}
\begin{array}{rcl}
-u_\rr{e}
&\!\!=&\!\!
 J^0_\bullet m_\rr{e}+2(J^0_{\circ_1}+J^0_{\circ_2})m_\rr{o}
 +2(J^0_{\langle\rangle_1}+J^0_{\langle\rangle_2})m_\rr{e}m_\rr{o}
 +(4J^0_{\langle\langle\rangle\rangle}
   +J^0_{\langle\langle\langle\rangle\rangle\rangle_1}
   +J^0_{\langle\langle\langle\rangle\rangle\rangle_2})
  m_\rr{o}^2
\\
&&\!\!
 +2(J^0_{\triangle_1}+J^0_{\triangle_2})m_\rr{o}^3
 +(4J^0_\llcorner+J^0_{\AC_1}+J^0_{\AC_2})m_\rr{e}m_\rr{o}^2
\\
&&\!\!
+2(J^0_{\perp_1}+J^0_{\perp_2})m_\rr{e}m_\rr{o}^3
+J^0_\diamond m_\rr{o}^4
+J^0_+ m_\rr{e}m_\rr{o}^4
\\
\end{array}
\end{equation}
A similar expression for $u_\rr{o}$ can be obtained by simply exchanging in 
\eqref{mf090} the role of~$\rr{e}$ and~$\rr{o}$. 

\subsection{The Mean Field equations}
\label{s:mfeq}
\par\noindent
By using \eqref{mf010}, \eqref{mf070}, \eqref{mf080}, and \eqref{mf090} 
we can finally write explicitly the free energy of the model in 
the Mean Field approximation. By minimizing such a free energy 
with respect to the magnetizations $m_\rr{e}$ and $m_\rr{o}$, 
namely, 
by setting $\partial f/\partial m_\rr{e}=\partial f/\partial m_\rr{o}=0$, 
with some algebra we get the two Mean Field equations 
\begin{equation}
\label{mf100}
m_\rr{r}
=
-\tanh\Big[\beta
         \frac{\partial}{\partial m_\rr{r}}(u_\rr{e}+u_\rr{o})
      \Big]
\;\;\;\textrm{ with }\;\;\;
\rr{r}=\rr{e},\rr{o}
\end{equation}
A straightforward computation gives the derivatives 
\begin{equation}
\label{mf110}
\begin{array}{rcl}
{\displaystyle
-\frac{\partial u_\rr{e}}{\partial m_\rr{e}}
}
&\!\!=&\!\!
J^0_\bullet
+2(J^0_{\langle\rangle_1}+J^0_{\langle\rangle_2})m_\rr{o}
+(4J^0_\llcorner+J^0_{\AC_1}+J^0_{\AC_2})m_\rr{o}^2
+2(J^0_{\perp_1}+J^0_{\perp_2})m_\rr{o}^3
\\
&&\!\!
+J^0_+ m_\rr{o}^4
\vphantom{\bigg\{_\big\}}
\\
{\displaystyle 
 -\frac{\partial u_\rr{e}}{\partial m_\rr{o}}
}
&\!\!=&\!\!
 2(J^0_{\circ_1}+J^0_{\circ_2})
 +2(J^0_{\langle\rangle_1}+J^0_{\langle\rangle_2})m_\rr{e}
 +2(4J^0_{\langle\langle\rangle\rangle}
   +J^0_{\langle\langle\langle\rangle\rangle\rangle_1}
   +J^0_{\langle\langle\langle\rangle\rangle\rangle_2})
  m_\rr{o}
\\
&&\!\!
 +6(J^0_{\triangle_1}+J^0_{\triangle_2})m_\rr{o}^2
 +2(4J^0_\llcorner+J^0_{\AC_1}+J^0_{\AC_2})m_\rr{e}m_\rr{o}
\\
&&\!\!
+6(J^0_{\perp_1}+J^0_{\perp_2})m_\rr{e}m_\rr{o}^2
+4J^0_\diamond m_\rr{o}^3
+4J^0_+ m_\rr{e}m_\rr{o}^3
\vphantom{\bigg\{_\big\}}
\\
\end{array}
\end{equation}
and
\begin{equation}
\label{mf120}
\begin{array}{rcl}
{\displaystyle 
 -\frac{\partial u_\rr{o}}{\partial m_\rr{e}}
}
&\!\!=&\!\!
 2(J^0_{\circ_1}+J^0_{\circ_2})
 +2(J^0_{\langle\rangle_1}+J^0_{\langle\rangle_2})m_\rr{o}
 +2(4J^0_{\langle\langle\rangle\rangle}
   +J^0_{\langle\langle\langle\rangle\rangle\rangle_1}
   +J^0_{\langle\langle\langle\rangle\rangle\rangle_2})
  m_\rr{e}
\\
&&\!\!
 +6(J^0_{\triangle_1}+J^0_{\triangle_2})m_\rr{e}^2
 +2(4J^0_\llcorner+J^0_{\AC_1}+J^0_{\AC_2})m_\rr{o}m_\rr{e}
\\
&&\!\!
+6(J^0_{\perp_1}+J^0_{\perp_2})m_\rr{o}m_\rr{e}^2
+4J^0_\diamond m_\rr{e}^3
+4J^0_+ m_\rr{o}m_\rr{e}^3
\vphantom{\bigg\{_\big\}}
\\
{\displaystyle
-\frac{\partial u_\rr{o}}{\partial m_\rr{o}}
}
&\!\!=&\!\!
J^0_\bullet
+2(J^0_{\langle\rangle_1}+J^0_{\langle\rangle_2})m_\rr{e}
+(4J^0_\llcorner+J^0_{\AC_1}+J^0_{\AC_2})m_\rr{e}^2
+2(J^0_{\perp_1}+J^0_{\perp_2})m_\rr{e}^3
\\
&&\!\!
+J^0_+ m_\rr{e}^4
\\
\end{array}
\end{equation}

\begin{figure}[t]
\begin{picture}(100,160)(-100,0)
\includegraphics[height=5.cm,angle=0]{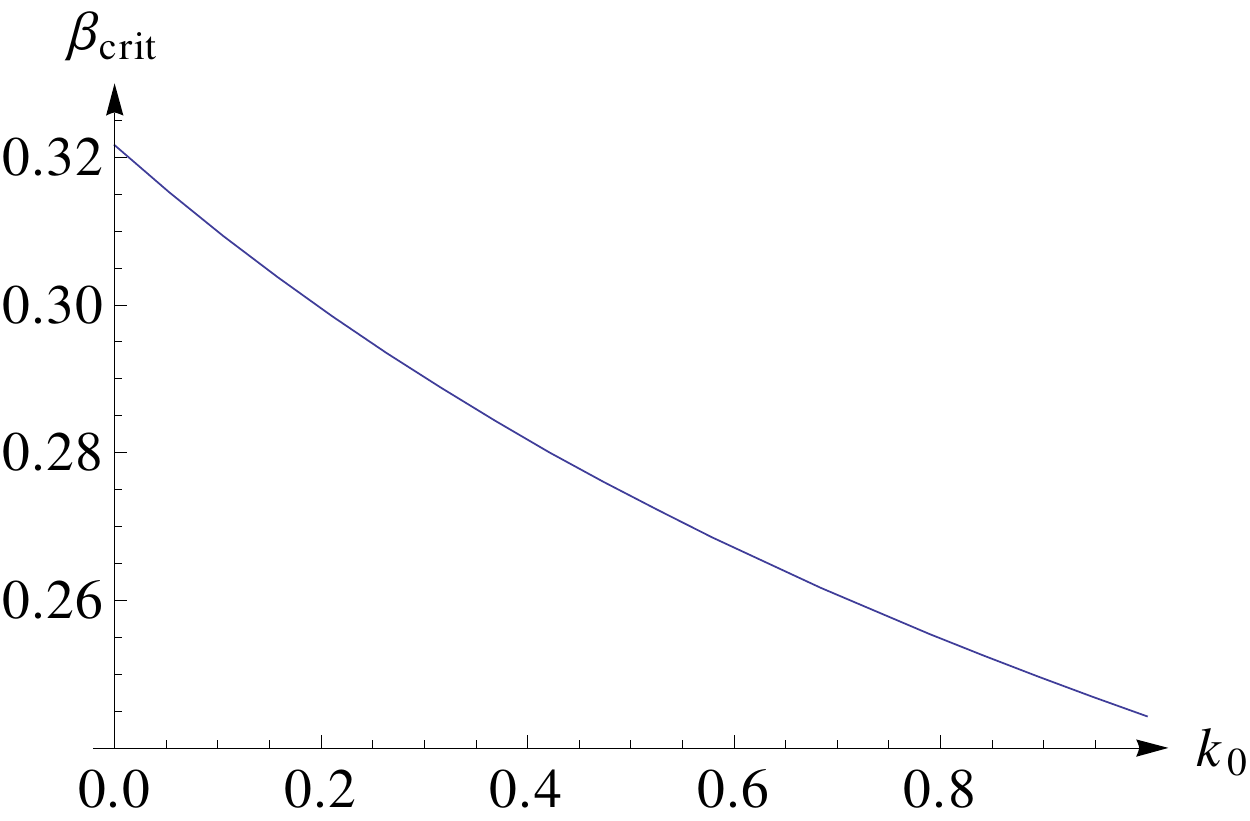}
\end{picture}
\caption{Critical temperature $\beta_\textrm{crit}$ as function 
of $k_0$ at $h=0$ and $k_1=k_2=1$.}
\label{f:mfres01}
\end{figure}

\subsection{Mean Field results}
\label{s:mfres}
\par\noindent
We are interested to study the behavior of the critical transition between the 
paramagnetic and the ferromagnetic one at $h=0$ and the structure 
of the low--temperature phase diagram. 

As discussed in Subsection~\ref{s:phase}, at $k_0=0$ the model 
behaves as two decoupled standard Ising models so that it exhibits 
a second order phase transition at $\beta=\log(1+\sqrt{2})/2\approx0.4407$
(the classical Onsager critical temperature).
By studying in this region the Mean Field equation, we find a 
continuous transition at $\beta=0.3216(5)$. 
In the case $k_1=k_2=1$ and $k_0=1$, it was shown in~\cite[section~7.2.3]{ThesePyl} through
Monte-Carlo simulations that  $\beta_{\textrm{crit}}\approx0.32$ to be compare with  $\beta_{\textrm{crit}}\approx0.24$ 
in the Mean Field approximation.
These results are not surprising at all, indeed the Mean Field approximation 
always provides a larger estimate for the critical temperature. 

What we are interested in is trying to understand the effect of the 
self-interaction on the behavior of the model. Hence, to understand if at $k_0>0$ the critical transition 
is still present and to estimate the value of the corresponding 
critical temperature. 

By solving the Mean Field equation for $k_0\in[0,1]$, we always 
find a critical transition and the value of the critical inverse 
temperature $\beta_\textrm{crit}(k_0)$ decreases when $k_0$ is increased.
Our results are depicted in figure~\ref{f:mfres01}.

\begin{figure}[t]
\begin{picture}(100,160)(-100,0)
\includegraphics[height=5.cm,angle=0]{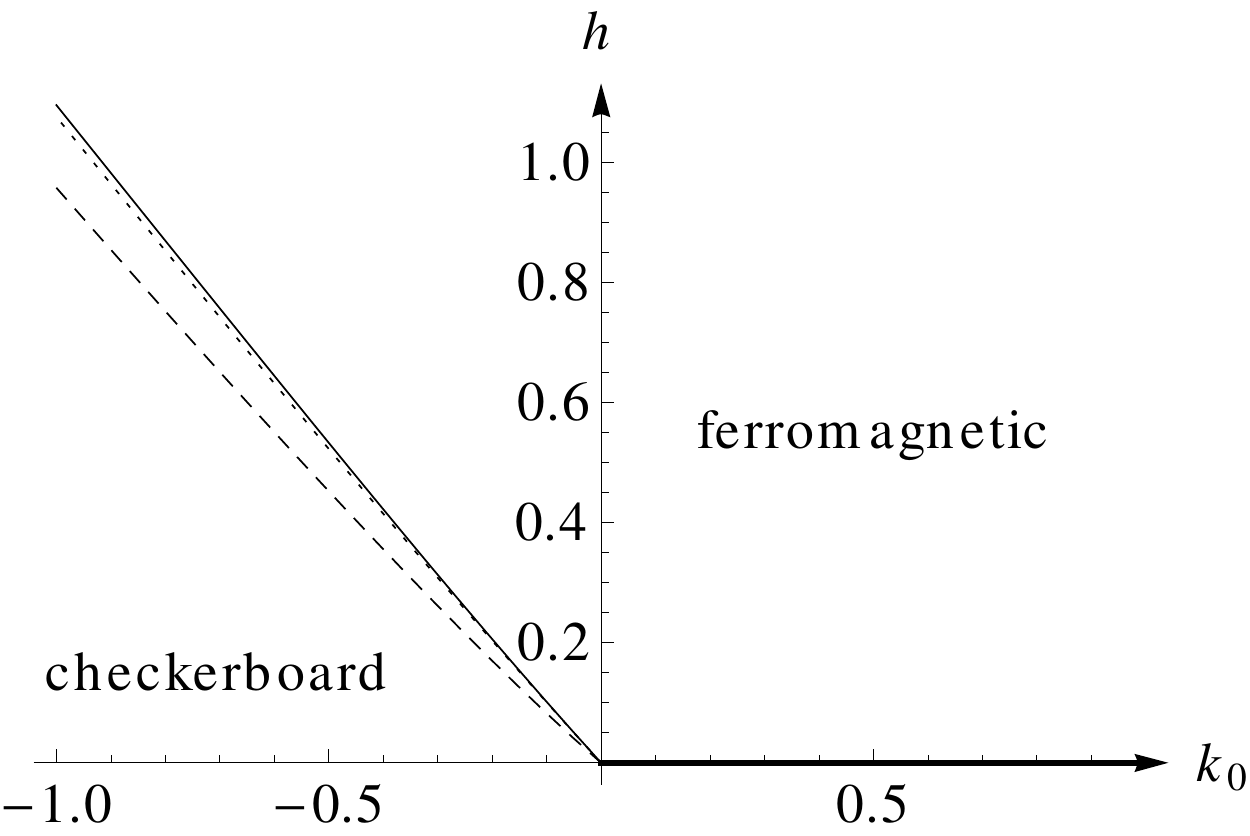}
\end{picture}
\caption{Phase diagram at different values of $\beta$ in the plane 
$k_0$--$h$ with $k_1=k_2=1$. 
In the region $k_0<0$ the three lines are the first order transition line 
between the positive ferromagnetic and the checkerboard phases. 
Continuous, dotted, and dashed lines refer, respectively, 
to the cases $\beta=1.2, 0.8, 0.4$. 
In the region $k_0>0$ the horizontal $h=0$ axis is the first order 
transition line between the two ferromagnetic phases.  
The phase digram in the region $h<0$ can be constructed by symmetry.}
\label{f:mfres02}
\end{figure}

As explained in the introduction, our main concern is that of 
understanding how the zero--temperature phase diagram, see 
figure~\ref{f:ground}, is modified at small positive temperatures. 
In view of the ground state structure we expect the presence 
of four phases: positive ferromagnetic, negative ferromagnetic, 
odd, and even checkerboard. The first two phases are respectively characterized
by positive and negative magnetization, while the last two by a 
positive and negative staggered magnetization. 

In the sequel we will not distinguish between the two checkerboard 
phases, since they are equivalent. Moreover, we shall discuss the phase 
diagram only in the region $h>0$. 
The case $h<0$ can then be recovered by using the spin--flip symmetry.  

For $h>0$, 
in order to draw the Mean Field phase diagram we identify 
the two different phases (checkerboard) and positive ferromagnetic 
by solving iteratively the Mean Field equations \eqref{mf100} starting 
from a suitable initial point, namely, $m_\rr{e}=m_\rr{o}=0.8$ 
for the ferromagnetic phase and 
$m_\rr{e}=0.8$ and $m_\rr{o}=-0.8$ 
for the checkerboard one. We shall decide about the phase of the 
system by choosing the one with smallest Mean Field free energy. The 
free energy will be computed by using 
\eqref{mf010}, \eqref{mf070}, \eqref{mf080}, and \eqref{mf090}. 

Mean Field predictions are summarized in figure~\ref{f:mfres02}, where 
the phase diagram of the cross PCA is plotted on the $k_0$--$h$ 
plane at different values of $\beta$. More precisely, 
we considered the values $\beta=1.2,0.8,0.4$. 

The most important remark is that the triple point is not 
affected by the temperature, indeed, its position is 
constantly the origin of the $k_0$--$h$ plane for each value of 
$\beta$. In other words the Mean Field 
approximation confirms the conjecture based on the entropy 
argument quoted in the introduction~\cite{CN}.

For the sake of completeness we briefly recall this argument.
At finite temperature, ground states are perturbed because small
droplets of different phases show up. The idea is to calculate the
energetic cost of a perturbation of one of the four
coexisting states via the formation of a square droplet of a
different phase. If it results that one of the four ground states
is more easily perturbed, then we will
conclude that this is the equilibrium phase at finite temperature.

The energy cost
of a square droplet of side length $\ell$ of one of the two
homogeneous ferromagnetic ground states plunged in one of the two 
checkerboards
(or vice versa) is equal to $8\ell$. On the other hand if an
homogeneous phase is perturbed as above by the other homogeneous
phases, or one of the two checkerboards is perturbed by the other
one, then the energy cost is $16\ell$.
Hence, from the energetic point of view the most convenient
excitations are those in which a homogeneous phase is perturbed
by a checkerboard or vice versa. Moreover, for each state
$\muno,\puno,\chess_\textrm{e},\chess_\textrm{o}$ 
there exist two possible energetically
convenient excitations: there is no entropic reason to prefer one
of the four ground states to the others when a finite 
low--temperature is considered. This is why it is possible 
to conjecture that at
small finite temperature the four ground states still coexist.

Finally we note that the Mean Field prediction for the 
ferro--checkerboard phase transition is that such a transition 
is discontinuous. And this result does not depend on the 
value of the temperature. 

\section{Conclusions}
\label{s:conclusioni}
\par\noindent
In this paper we have discussed some general properties 
of the Hamiltonian associated with 
a class of reversible Probabilistic Cellular Automata. 
We have focused our attention to the so--called cross PCA model, 
which is a two--dimensional 
reversible PCA in which the updating rule of a cell 
depends on the status of the five cells forming a cross 
centered at the cell itself. 

This model had been extensively studied from the metastability 
point of view and many interesting properties have been shown. 
In particular a suggestive analogy with the Blume--Capel model 
had been pointed out in the metastability literature. 

In this paper we focused our attention on the structure of the 
potentials describing the microscopic interaction and on 
the zero and positive small temperature phase diagram. 
We computed the zero--temperature phase diagram exactly with respect to 
the self--interaction intensity and the magnetic field. 

At finite temperature the phase diagram has been derived in the 
framework of a suitable Mean Field approximation. We have 
discussed the variation of the critical temperature for the 
transition between the ordered and the disordered phase 
at zero magnetic field as a function of the self--interaction intensity. 
We have shown that, in the Mean Field approximation, such a temperature 
is an increasing function of the self--interaction intensity. 

Moreover we have discussed the low--temperature phase diagram in the 
plane $k_0$--$h$ and have shown that the topology of the 
zero temperature phase diagram is preserved when the temperature 
is positive and small. Finally, we have shown that the Mean Field 
approximation is consistent with an entropic heuristic argument 
suggesting that the position of the zero--temperature triple point 
is not changed at low--temperature. 

\appendix

\appendice{Coupling constants}
\label{s:accoppiamenti}
\par\noindent
In this appendix we report the expression of the coupling constants 
defined in Section~\ref{s:ham-croce} (see also figure~\ref{f:accoppiamenti})
as function of $h$, $\beta$, $k_0$, $k_1$, and $k_2$. The scheme 
we adopted for the computation is described in Section~\ref{s:ham-croce} 
as well.
For the couplings in which we had to distinguish between the horizontal 
and the vertical case we report only the horizontal one and note that 
the corresponding vertical one can be obtained by exchanging in the 
formula the role of $k_1$ and $k_2$. 

\begin{equation} %
\begin{split}
J^0_{\bullet}
&=\frac{1}{5}h+\frac{1}{2^5\beta}\log\bigg\{
     \frac
          {
           \cosh^4[\beta(h+k_0)]
          }
          {
           \cosh^4[\beta(h-k_0)]
          }
\\
&
     \phantom{h+\frac{1}{2^5\beta}\log\bigg\{\times}
     \times
     \frac
          {
           \cosh^2[\beta(h+k_0+2k_1)]
           \cosh^2[\beta(h+k_0-2k_1)]
          }
          {
           \cosh^2[\beta(h-k_0+2k_1)]
           \cosh^2[\beta(h-k_0-2k_1)]
          }
\\
&
     \phantom{h+\frac{1}{2^5\beta}\log\bigg\{\times}
     \times
     \frac
          {
           \cosh^2[\beta(h+k_0+2k_2)]
           \cosh^2[\beta(h+k_0-2k_2)]
          }
          {
           \cosh^2[\beta(h-k_0+2k_2)]
           \cosh^2[\beta(h-k_0-2k_2)]
          }
\\
&
     \phantom{h+\frac{1}{2^5\beta}\log\bigg\{\times}
     \times
     \frac
          {
           \cosh[\beta(h+k_0+2(k_1-k_2))]
           \cosh[\beta(h+k_0-2(k_1-k_2))]
          }
          {
           \cosh[\beta(h-k_0+2(k_1-k_2))]
           \cosh[\beta(h-k_0-2(k_1-k_2))]
          }
\\
&
     \phantom{h+\frac{1}{2^5\beta}\log\bigg\{\times}
     \times
     \frac
          {
           \cosh[\beta(h+k_0+2(k_1+k_2))]
           \cosh[\beta(h+k_0-2(k_1+k_2))]
          }
          {
           \cosh[\beta(h-k_0+2(k_1+k_2))]
           \cosh[\beta(h-k_0-2(k_1+k_2))]
          }
\bigg\}
\end{split}
\end{equation}

\begin{equation} %
\begin{split}
J^0_{\circ_1}
&=\frac{1}{5}h+\frac{1}{2^5\beta}\log\bigg\{
     \frac
          {
           \cosh^2[\beta(h-k_0+2k_1)]
           \cosh^2[\beta(h+k_0+2k_1)]
          }
          {
           \cosh^2[\beta(h-k_0-2k_1)]
           \cosh^2[\beta(h+k_0-2k_1)]
          }
\\
&
     \phantom{h+\frac{1}{2^5\beta}\log\bigg\{\times}
     \times
     \frac
          {
           \cosh[\beta(h+k_0+2(k_1-k_2))]
           \cosh[\beta(h-k_0+2(k_1-k_2))]
          }
          {
           \cosh[\beta(h+k_0-2(k_1-k_2))]
           \cosh[\beta(h-k_0-2(k_1-k_2))]
          }
\\
&
     \phantom{h+\frac{1}{2^5\beta}\log\bigg\{\times}
     \times
     \frac
          {
           \cosh[\beta(h+k_0+2(k_1+k_2))]
           \cosh[\beta(h-k_0+2(k_1+k_2))]
          }
          {
           \cosh[\beta(h+k_0-2(k_1+k_2))]
           \cosh[\beta(h-k_0-2(k_1+k_2))]
          }
\bigg\}
\end{split}
\end{equation}


\begin{equation} %
\begin{split}
J^0_{\langle\rangle_1}
&=\frac{1}{2^5\beta}\log\bigg\{
     \frac
          {
           \cosh^2[\beta(h-k_0-2k_1)]
           \cosh^2[\beta(h+k_0+2k_1)]
          }
          {
           \cosh^2[\beta(h+k_0-2k_1)]
           \cosh^2[\beta(h-k_0+2k_1)]
          }
\\
&
     \phantom{\frac{1}{2^5\beta}\log\bigg\{\times}
     \times
     \frac
          {
           \cosh[\beta(h+k_0+2k_1-2k_2)]
           \cosh[\beta(h-k_0-2k_1+2k_2)]
          }
          {
           \cosh[\beta(h-k_0+2k_1-2k_2)]
           \cosh[\beta(h+k_0-2k_1+2k_2)]
          }
\\
&
     \phantom{\frac{1}{2^5\beta}\log\bigg\{\times}
     \times
     \frac
          {
           \cosh[\beta(h-k_0-2(k_1+k_2))]
           \cosh[\beta(h+k_0+2(k_1+k_2))]
          }
          {
           \cosh[\beta(h+k_0-2(k_1+k_2))]
           \cosh[\beta(h-k_0+2(k_1+k_2))]
          }
\bigg\}
\end{split}
\end{equation}


\begin{equation} %
\begin{split}
J^0_{\langle\langle\rangle\rangle}
&=\frac{1}{2^5\beta}\log\bigg\{
     \frac
          {
           \cosh[\beta(h-k_0-2(k_1+k_2))]
           \cosh[\beta(h+k_0-2(k_1+k_2))]
          }
          {
           \cosh[\beta(h-k_0+2k_1-2k_2)]
           \cosh[\beta(h+k_0+2k_1-2k_2)]
          }
\\
&
     \phantom{\frac{1}{2^5\beta}\log\bigg\{\times}
     \times
     \frac
          {
           \cosh[\beta(h-k_0+2(k_1+k_2))]
           \cosh[\beta(h+k_0+2(k_1+k_2))]
          }
          {
           \cosh[\beta(h-k_0-2k_1+2k_2)]
           \cosh[\beta(h+k_0-2k_1+2k_2)]
          }
\bigg\}
\end{split}
\end{equation}

\begin{equation} %
\begin{split}
J^0_{\langle\langle\langle\rangle\rangle\rangle_1}
&=\frac{1}{2^5\beta}\log\bigg\{
     \frac
          {
           \cosh^2[\beta(h-k_0-2k_1)]
           \cosh^2[\beta(h+k_0-2k_1)]
          }
          {
           \cosh^2[\beta(h-k_0-2k_2)]
           \cosh^2[\beta(h+k_0-2k_2)]
          }
\\
&
     \phantom{\frac{1}{2^5\beta}\log\bigg\{\times}
     \times
     \frac
          {
           \cosh^2[\beta(h-k_0+2k_1)]
           \cosh^2[\beta(h+k_0+2k_1)]
          }
          {
           \cosh^2[\beta(h-k_0+2k_2)]
           \cosh^2[\beta(h+k_0+2k_2)]
          }
\\
&
     \phantom{\frac{1}{2^5\beta}\log\bigg\{\times}
     \times
     \frac
          {1
          }
          {
           \cosh^4[\beta(h+k_0)]
           \cosh^4[\beta(h-k_0)]
          }
\\
&
     \phantom{\frac{1}{2^5\beta}\log\bigg\{\times}
     \times
           \cosh[\beta(h-k_0+2(k_1-k_2))]
           \cosh[\beta(h+k_0+2(k_1-k_2))]
\\
&
     \phantom{\frac{1}{2^5\beta}\log\bigg\{\times}
     \times
           \cosh[\beta(h-k_0-2(k_1-k_2))]
           \cosh[\beta(h+k_0-2(k_1-k_2))]
\\
&
     \phantom{\frac{1}{2^5\beta}\log\bigg\{\times}
     \times
           \cosh[\beta(h-k_0-2(k_1+k_2))]
           \cosh[\beta(h+k_0-2(k_1+k_2))]
\\
&
     \phantom{\frac{1}{2^5\beta}\log\bigg\{\times}
     \times
           \cosh[\beta(h-k_0+2(k_1+k_2))]
           \cosh[\beta(h+k_0+2(k_1+k_2))]
\bigg\}
\end{split}
\end{equation}


\begin{equation} %
\begin{split}
J^0_{\triangle_1}
&=\frac{1}{2^5\beta}\log\bigg\{
     \frac
          {
           \cosh^2[\beta(h-k_0-2k_2)]
           \cosh^2[\beta(h+k_0-2k_2)]
          }
          {
           \cosh^2[\beta(h-k_0+2k_2)]
           \cosh^2[\beta(h+k_0+2k_2)]
          }
\\
&
     \phantom{\frac{1}{2^5\beta}\log\bigg\{\times}
     \times
     \frac
          {
           \cosh[\beta(h-k_0-2k_1+2k_2)]
           \cosh[\beta(h+k_0-2k_1+2k_2)]
          }
          {
           \cosh[\beta(h-k_0+2k_1-2k_2)]
           \cosh[\beta(h+k_0+2k_1-2k_2)]
          }
\\
&
     \phantom{\frac{1}{2^5\beta}\log\bigg\{\times}
     \times
     \frac
          {
           \cosh[\beta(h-k_0+2(k_1+k_2))]
           \cosh[\beta(h+k_0+2(k_1+k_2))]
          }
          {
           \cosh[\beta(h-k_0-2(k_1+k_2))]
           \cosh[\beta(h+k_0-2(k_1+k_2))]
          }
\bigg\}
\end{split}
\end{equation}


\begin{equation} %
\begin{split}
J^0_{\llcorner}
&=\frac{1}{2^5\beta}\log\bigg\{
     \frac
          {
           \cosh[\beta(h-k_0+2(k_1-k_2))]
           \cosh[\beta(h-k_0-2(k_1-k_2))]
          }
          {
           \cosh[\beta(h+k_0+2(k_1-k_2))]
           \cosh[\beta(h+k_0-2(k_1-k_2))]
          }
\\
&
     \phantom{\frac{1}{2^5\beta}\log\bigg\{\times}
     \times
     \frac
          {
           \cosh[\beta(h+k_0-2(k_1+k_2))]
           \cosh[\beta(h+k_0+2(k_1+k_2))]
          }
          {
           \cosh[\beta(h-k_0-2(k_1+k_2))]
           \cosh[\beta(h-k_0+2(k_1+k_2))]
          }
\bigg\}
\end{split}
\end{equation}

\begin{equation} %
\begin{split}
J^0_{\AC_1}
&=\frac{1}{2^5\beta}\log\bigg\{
     \frac
          {
           \cosh^4[\beta(h-k_0)]
           \cosh^2[\beta(h+k_0-2k_1)]
           \cosh^2[\beta(h+k_0+2k_1)]
          }
          {
           \cosh^4[\beta(h+k_0)]
           \cosh^2[\beta(h-k_0-2k_1)]
           \cosh^2[\beta(h-k_0+2k_1)]
          }
\\
&
     \phantom{\frac{1}{2^5\beta}\log\bigg\{\times}
     \times
     \frac
          {
           \cosh^2[\beta(h-k_0-2k_2)]
           \cosh^2[\beta(h-k_0+2k_2)]
          }
          {
           \cosh^2[\beta(h+k_0-2k_2)]
           \cosh^2[\beta(h+k_0+2k_2)]
          }
\\
&
     \phantom{\frac{1}{2^5\beta}\log\bigg\{\times}
     \times
     \frac
          {
           \cosh[\beta(h+k_0+2(k_1-k_2))]
           \cosh[\beta(h+k_0-2(k_1-k_2))]
          }
          {
           \cosh[\beta(h-k_0+2(k_1-k_2))]
           \cosh[\beta(h-k_0-2(k_1-k_2))]
          }
\\
&
     \phantom{\frac{1}{2^5\beta}\log\bigg\{\times}
     \times
     \frac
          {
           \cosh[\beta(h+k_0-2(k_1+k_2))]
           \cosh[\beta(h+k_0+2(k_1+k_2))]
          }
          {
           \cosh[\beta(h-k_0-2(k_1+k_2))]
           \cosh[\beta(h-k_0+2(k_1+k_2))]
          }
\bigg\}
\end{split}
\end{equation}


\begin{equation} %
\begin{split}
J^0_{\perp_1}
&=\frac{1}{2^5\beta}\log\bigg\{
     \frac
          {
           \cosh^2[\beta(h+k_0-2k_2)]
           \cosh^2[\beta(h-k_0+2k_2)]
          }
          {
           \cosh^2[\beta(h-k_0-2k_2)]
           \cosh^2[\beta(h+k_0+2k_2)]
          }
\\
&
     \phantom{\frac{1}{2^5\beta}\log\bigg\{\times}
     \times
     \frac
          {
           \cosh[\beta(h-k_0+2(k_1-k_2))]
           \cosh[\beta(h+k_0-2(k_1-k_2))]
          }
          {
           \cosh[\beta(h+k_0+2(k_1-k_2))]
           \cosh[\beta(h-k_0-2(k_1-k_2))]
          }
\\
&
     \phantom{\frac{1}{2^5\beta}\log\bigg\{\times}
     \times
     \frac
          {
           \cosh[\beta(h-k_0-2(k_1+k_2))]
           \cosh[\beta(h+k_0+2(k_1+k_2))]
          }
          {
           \cosh[\beta(h+k_0-2(k_1+k_2))]
           \cosh[\beta(h-k_0+2(k_1+k_2))]
          }
\bigg\}
\end{split}
\end{equation}


\begin{equation} %
\begin{split}
J^0_{\diamondsuit}
&=\frac{1}{2^5\beta}\log\bigg\{
           \cosh^4[\beta(h-k_0)]
           \cosh^4[\beta(h+k_0)]
\\
&
     \phantom{\frac{1}{2^5\beta}\log\bigg\{\times}
     \times
     \frac
          {
           \cosh[\beta(h-k_0-2(k_1+k_2))]
           \cosh[\beta(h+k_0-2(k_1+k_2))]
          }
          {
           \cosh^2[\beta(h-k_0-2k_1)]
           \cosh^2[\beta(h+k_0-2k_1)]
          }
\\
&
     \phantom{\frac{1}{2^5\beta}\log\bigg\{\times}
     \times
     \frac
          {
           \cosh[\beta(h-k_0+2(k_1+k_2))]
           \cosh[\beta(h+k_0+2(k_1+k_2))]
          }
          {
           \cosh^2[\beta(h-k_0+2k_1)]
           \cosh^2[\beta(h+k_0+2k_1)]
          }
\\
&
     \phantom{\frac{1}{2^5\beta}\log\bigg\{\times}
     \times
     \frac
          {
           \cosh[\beta(h-k_0+2(k_1-k_2))]
           \cosh[\beta(h+k_0+2(k_1-k_2))]
          }
          {
           \cosh^2[\beta(h-k_0-2k_2)]
           \cosh^2[\beta(h+k_0-2k_2)]
          }
\\
&
     \phantom{\frac{1}{2^5\beta}\log\bigg\{\times}
     \times
     \frac
          {
           \cosh[\beta(h-k_0-2(k_1-k_2))]
           \cosh[\beta(h+k_0-2(k_1-k_2))]
          }
          {
           \cosh^2[\beta(h-k_0+2k_2)]
           \cosh^2[\beta(h+k_0+2k_2)]
          }
\bigg\}
\end{split}
\end{equation}

\begin{equation} %
\begin{split}
J^0_{+}
&=\frac{1}{2^5\beta}\log\bigg\{
     \frac
          {
           \cosh^4[\beta(h+k_0)]
           \cosh^2[\beta(h-k_0-2k_1)]
           \cosh^2[\beta(h-k_0+2k_1)]
          }
          {
           \cosh^4[\beta(h-k_0)]
           \cosh^2[\beta(h+k_0-2k_1)]
           \cosh^2[\beta(h+k_0+2k_1)]
          }
\\
&
     \phantom{\frac{1}{2^5\beta}\log\bigg\{\times}
     \times
     \frac
          {
           \cosh^2[\beta(h-k_0-2k_2)]
           \cosh^2[\beta(h-k_0+2k_2)]
          }
          {
           \cosh^2[\beta(h+k_0-2k_2)]
           \cosh^2[\beta(h+k_0+2k_2)]
          }
\\
&
     \phantom{\frac{1}{2^5\beta}\log\bigg\{\times}
     \times
     \frac
          {
           \cosh[\beta(h+k_0+2(k_1-k_2))]
           \cosh[\beta(h+k_0-2(k_1-k_2))]
          }
          {
           \cosh[\beta(h-k_0+2(k_1-k_2))]
           \cosh[\beta(h-k_0-2(k_1-k_2))]
          }
\\
&
     \phantom{\frac{1}{2^5\beta}\log\bigg\{\times}
     \times
     \frac
          {
           \cosh[\beta(h+k_0-2(k_1+k_2))]
           \cosh[\beta(h+k_0+2(k_1+k_2))]
          }
          {
           \cosh[\beta(h-k_0-2(k_1+k_2))]
           \cosh[\beta(h-k_0+2(k_1+k_2))]
          }
\bigg\}
\end{split}
\end{equation}

\bigskip
\par\noindent
\textbf{Acknowledgments.\/} 
The authors thank J.\ Bricmont, A.\ Pelizzola, 
and A.\ van Enter for some very useful discussions, comments, and 
references.
P.--Y. Louis thanks EURANDOM/TU Eindhoven, Utrecht University, 
and Dipartimento SBAI (Sapienza Universit\`a di Roma)
where part of this work was done and the CNRS for supporting these 
research stays.
E.N.M.\ Cirillo thanks Technical University Delft, Utrecht Mathematics Department 
and EURANDOM/TU Eindhoven for their kind hospitality and financial support.


\begin{thebibliography}{99}

\bibitem{BCLS}
S.\ Bigelis, E.N.M.\ Cirillo, J.L.\ Lebowitz, E.R.\ Speer,
``Critical droplets in metastable probabilistic cellular automata,"
{\sl Phys.\ Rev.\ E} \textbf{59}, 3935, (1999).

\bibitem{B}
M.\ Blume,
``Theory of the First--Order Magnetic Phase Change in UO$_2$."
{\sl Phys.\ Rev.\/} \textbf{141}, 517, (1966).

\bibitem{BEG} M.\ Blume, V.J.\ Emery, R.B.\ Griffiths, 
``Ising Model for the $\lambda$ Transition and 
  Phase Separation in He$^3$--He$^4$ Mixtures."
{\sl Phys.\ Rev.\ A} \textbf{4}, 1071--1077, (1971).

\bibitem{BS}
J.\ Bricmont, J.\ Slawny, 
``Phase Transitions in Systems with a Finite Number of Dominant 
Ground States.''
{\sl Journ.\ Stat.\ Phys.\/} \textbf{54}, 89, (1989).

\bibitem{C}
H.W.\ Capel,
``On possibility of first--order phase transitions in Ising systems of 
  triplet ions with zero--field splitting."
{\sl Physica} \textbf{32}, 966, (1966).

\bibitem{CCGM}
A.\ Cappi, P.\ Colangelo, G.\ Gonnella, A.\ Maritan,
``Ensemble of interacting randon surfaces on a lattice."
{\sl Nuclear Physics B} \textbf{370}, 659--694, (1992).

\bibitem{CNS1}
E.N.M.\ Cirillo, F.R.\ Nardi, C.\ Spitoni,
``Metastability for a reversible probabilistic cellular
automata with self--interaction."
{\sl Journ.\ Stat.\ Phys.} {\bf 132}, 431--471, (2008).

\bibitem{CN}
E.N.M.\ Cirillo, F,R.\ Nardi,
``Metastability for the Ising model with a parallel dynamics."
{\sl Journ.\ Stat.\ Phys.\/} {\bf 110}, 183--217, (2003).

\bibitem{CNS2}
E.N.M.\ Cirillo, F.R.\ Nardi, C.\ Spitoni,
``Competitive nucleation in reversible Probabilistic
Cellular Automata."
{\sl Physical Review E} {\bf 78}, 040601, (2008).

\bibitem{CNS3}
E.N.M.\ Cirillo, F.R.\ Nardi, C.\ Spitoni,
``Competitive nucleation in metastable systems."
Applied and Industrial Mathematics in Italy III.
Series on Advances in Mathematics for Applied Sciences,
Volume {\bf 82}, 208--219, (2010).

\bibitem{CNP}
E.N.M.\ Cirillo, F.R.\ Nardi, A.D.\ Polosa,
``Magnetic order in the Ising model with parallel dynamics."
{\sl Phys.\ Rev.\ E} {\bf 64}, 57103, (2001).

\bibitem{CO}
E.N.M.\ Cirillo, E.\ Olivieri,
``Metastability and nucleation for the Blume-Capel model.
Different mechanisms of transition,''
{\sl Journ.\ Stat.\ Phys.\/} {\bf 83}, 473--554, (1996).

\bibitem{CS96}
E.N.M.\ Cirillo, S.\ Stramaglia,
``Polymerization in a ferromagnetic spin model with threshold."
{\sl Phys.\ Rev.\ E} {\bf 54}, 1096, (1996).

\bibitem{DpLR}
P.\ Dai Pra, P.-Y.\ Louis, S.\ Roelly,
``Stationary measures and phase transition for a class of probabilistic
cellular automata."
{\sl ESAIM Probab.\ Statist.\/}  {\bf 6}, 89--104, (2002).

\bibitem{Derrida}
B.\ Derrida,
``Dynamical phase transition in spin model and automata,"
Fundamental problem in Statistical Mechanics VII, H.\ van Beijeren, Editor,
Elsevier Science,  (1990).

\bibitem{GeorgesLeDoussal}
A.\ Georges and P.\ Le Doussal,
``From equilibrium spin models to probabilistic cellular automata",
{\sl Journal of Statistical Physics}, \textbf{54}, 3-4, 1011--1064, (1989).

\bibitem{gklm}
 Goldstein S., Kuik R., Lebowitz J.L., Maes C., 
``From PCAs to equilibrium systems and back",
{\sl Comm. Math. Phys.\/}, \textbf{125}, no. 1, 71--79, (1989). 

\bibitem{GJH}
G.\ Grinstein, C.\ Jayaprakash, and Y.\ He, 
``Statistical Mechanics of Probabilistic Cellular Automata"
{\sl Phys. Rev. Lett.\/} \textbf{55}, 2527, (1985).

\bibitem{HK}
K.\ Haller, T.\ Kennedy,
``Absence of renormalization group pathologies near the critical 
temperature. Two examples.''
{\sl Journ.\ Stat.\ Phys.\/} {\bf 85}, 607--637, (1996).


\bibitem{Kari}
J.\ Kari,
``Theory of cellular automata: A survey'',
{\sl Theoretical Computer Science}, Volume \textbf{334}, Issues 1-3, 3--33, (2005).

\bibitem{KV}
V.\ Kozlov, N.B.\ Vasiljev,
``Reversible Markov chain with local interactions," in
``Multicomponent random system," 451--469, Adv.\ in Prob.\ and Rel.\ Topics, (1980).

\bibitem{NonLinearBook}
L.\ Lam, 
``Non-Linear Physics for Beginners: Fractals, Chaos, Pattern Formation, 
Solitons, Cellular Automata and Complex Systems",
book, World Scientific, (1998).

\bibitem{lms}
 Lebowitz J.L., Maes C., Speer E.R., 
``{Statistical mechanics of probabilistic cellular
automata.}", {\sl J. Statist. Phys.} \textbf{59}, 1-2, 117--170, (1990).

\bibitem{ThesePyl}
P.-Y.\ Louis.
``{Automates Cellulaires Probabilistes : mesures stationnaires,
  mesures de Gibbs associ{\'e}es et ergodicit{\'e}}",
{\sl PhD thesis}, Universit{\'e} {Lille~I} and Politecnico di Milano, sept. 2002.
\url{http://tel.archives-ouvertes.fr/index.php?halsid=2amredt6j0va4869evh5q209e3&view_this_doc=tel-00002203&version=2}

\bibitem{EspTempsPyl}
P.-Y.\ Louis.
 ``Ergodicity of {PCA}: equivalence between spatial and temporal mixing
  conditions", 
 {\sl Electronic Communications in Probability}, \textbf{9}, 119--131, (2004).

\bibitem{LR}
P.-Y.\ Louis and J.-B.\ Rouquier.
``Time-to-coalescence for interacting particle systems: Parallel versus
  sequential updating.",
{\sl Preprint 2009/03}, Universit\"at Potsdam, ISSN~no.~1613-3307 (2009)\\
\url{http://nbn-resolving.de/urn:nbn:de:kobv:517-opus-49454}.

\bibitem{NS}
F.R.\ Nardi, C.\ Spitoni,
``Sharp asymptotics for stochastic dynamics with parallel updating rule."
{\sl Journal of Statistical Physics}, \textbf{146}, 701--718, (2012).

\bibitem{PCA:order:disorder}
 J.\ Palandi, R.M.C.\ de Almeida, J.R.\ Iglesias, M.\ Kiwi, 
``Cellular automaton for the order-disorder transition", 
{\sl Chaos, Solitons \& Fractals}, Vol. \textbf{6}, 439--445, (1995).

\bibitem{PGSFM}
M.\ Perc, J.\ G\'omez--Garde\~{n}es, A.\ Szolnoki, L.M.\ Flor\'ia, 
and Y.\ Moreno,
``Evolutionary dynamics of group interactions on structured populations: 
A review",
{\sl J.\ R.\ Soc.\ Interface} \textbf{10}, 20120997, (2013).

\bibitem{PG}
M.\ Perc and P.\ Grigolini, 
``Collective behavior and evolutionary games -- An introduction",
{\sl Chaos, Solitons \& Fractals} \textbf{56}, 1--5, (2013).
	
\bibitem{PS}
M.\ Perc and A.\ Szolnoki, 
``Coevolutionary games -- A mini review",
{\sl Biosystems} \textbf{99}, 109--125, (2010).

\bibitem{sirakoulis2012cellular}
G.Ch.\ Sirakoulis  and S.\ Bandini (editors),
``Cellular Automata: 10th International Conference on Cellular Automata for Research and Industry'',
 ACRI 2012, {\sl Proceedings, 2012, Lecture Notes in Computer Science}, Springer (2012).

\bibitem{S}
J.\ Slawny, 
``Low--Temperature Properties of Classical Lattice Systems: Phase 
Transitions and Phase Diagrams",
in \textit{Phase Transitions and Critical Phenomena}, 
vol.\ \textbf{11}, C.~Domb 
and J.L.~Lebowitz, eds.\ Academic Press, London, (1987).

\bibitem{tvs}
A.L.\ Toom , N.B.\ Vasilyev, O.N.\ Stavskaya, L.G.\ Mityushin,
G.L.\ Kurdyumov, S.A.\ Pirogov, {\sl Discrete local Markov systems},
in \textit{Stochastic Cellular Systems: ergodicity, memory, morphogenesis},
edited by R.L. Dobrushin, V.I. Kryukov, A.L. Toom, Manchester
University Press, 1--182, (1990).

\bibitem{Vas69}
L.N.\ Vasershtein, 
''Markov processes over denumerable products of spaces
describing large system of automata", 
{\sl Problemy Pereda\v{c}i Informacii} \textbf{5}, no. 3, 64--72, (1969).

\bibitem{Wolfram:RevModPhys.55.601}
S.\ Wolfram,
``Statistical mechanics of cellular automata",
  {\sl Rev. Mod. Phys.}, \textbf{55}, {3}, 601--644,
(1983).

\bibitem{Wolfram1984}
S.\ Wolfram, 
''Universality and complexity in cellular automata``,
{\sl Physica D: Nonlinear Phenomena}, Vol. \textbf{10}, Issues 1-2, 1--35, (1984).

\bibitem{Wolfram1984Nature}
S.\ Wolfram,
``Cellular automata as models of complexity",
{\sl Nature} \textbf{311}, 419--424, (1984).

\end{thebibliography}
\end{document}